\documentclass[prd,preprint,nofootinbib]{revtex4}

\usepackage{graphicx}
\usepackage{amssymb}
\usepackage{amsmath}
\usepackage{color}

\begin{document}
 
\newcommand{\be}{\begin{equation}}
\newcommand{\ee}{\end{equation}}
\newcommand{\aprime}{\mathbf{a}^{\prime}}
\newcommand{\bprime}{\mathbf{b}^{\prime}}
\newcommand{\kh}{\hat{k}}
\newcommand{\Ip}{\vec{I}_+}
\newcommand{\Imi}{\vec{I}_-}
\newcommand{\bc}{\begin{cases}}
\newcommand{\ec}{\end{cases}}

\newcommand{\red}{\color{red}}
\newcommand{\cyan}{\color{cyan}}
\newcommand{\blue}{\color{blue}}
\newcommand{\magenta}{\color{magenta}}
\newcommand{\yellow}{\color{yellow}}
\newcommand{\green}{\color{green}}
\newcommand{\rem}[1]{{\bf\blue #1}}

\begin{flushright}
ICRR-Report-558 \\
IPMU 10-0017 \\
\end{flushright}

%\vskip 1cm

\title{
Gravitational waves from kinks on infinite cosmic strings
}

%\vskip 1cm

\author{Masahiro Kawasaki$^{(a, b)}$, Koichi Miyamoto$^{(a)}$ 
and Kazunori Nakayama$^{(a)}$}

\affiliation{%
$^a$Institute for Cosmic Ray Research,
     University of Tokyo, Kashiwa, Chiba 277-8582, Japan\\
$^b$Institute for the Physics and Mathematics of the Universe, 
     University of Tokyo, Kashiwa, Chiba 277-8568, Japan
}

\date{\today}

\vskip 1.0cm

\begin{abstract}
Gravitational waves emitted by kinks on infinite strings are
investigated using detailed estimations of the kink distribution on
infinite strings.  We find that gravitational waves from kinks can be
detected by future pulsar timing experiments such as SKA for an
appropriate value of the the string tension, if the typical size of
string loops is much smaller than the horizon at their formation.
Moreover, the gravitational wave spectrum depends on the thermal history
of the Universe and hence it can be used as a probe into the early
evolution of the Universe.
\end{abstract}

 \maketitle

%%%%%%%%%%%%%%%%%%%%%%%%%%%%%%%%%%%%%%%%%%%%%% 
 \section{Introduction}
 %%%%%%%%%%%%%%%%%%%%%%%%%%%%%%%%%%%%%%%%%%%%%%
 
It is well known that cosmic strings are produced in the early Universe
at the phase transition associated with spontaneous symmetry breaking in
the Grand-Unified-Theories (GUTs)~\cite{Vilenkin}. Although cosmic strings 
formed before inflation are diluted away, some GUTs predict series 
of phase transitions where symmetry such as $U(1)_{B-L}$  ($B$ and $L$ 
denotes baryon and lepton number)  breaks at low energy and produces 
strings after inflation.  
Cosmic strings are also produced at the end of brane inflation
in the framework of the superstring theory~\cite{Sarangi:2002yt,Dvali:2003zj}. 
Once produced, they
survive until now and can  leave observable signatures.  Thus, the cosmic
strings provide us with an opportunity to probe unified theories in
particle physics which cannot be tested in terrestrial experiments.

Various cosmological and astrophysical signals of cosmic strings have
been intensively studied for decades. Especially, many authors have
studied gravitational waves (GW) emitted from the cosmic string network,
in particular, GWs from cosmic string loops.  Cosmic string loops
oscillate by their tension and emit low frequency GWs corresponding to
their size~\cite{Caldwell:1991jj,DePies:2007bm}. Moreover, string loops
generically have cusps and kinks, and these structures cause high
frequency modes of GWs, i.e. GW bursts~%
\cite{Damour:2000wa,Damour:2001bk,Damour:2004kw,Siemens:2006yp}.

On the other hand, GWs from infinite strings (long strings which lie
across the Hubble horizon) have attracted much less interest than those
from loops.  This is because there are only a few infinite strings
in one Hubble horizon, so GWs from them are much weaker than those from
loops whose number density in the Hubble volume is much larger.\footnote{%%
For cosmic strings whose reconnection probability is small, such as
cosmic superstrings, many infinite strings can exist in a Hubble
horizon. For such kinds of cosmic strings, however, the number of loops
in a horizon is accordingly large, hence the situation that the loop
contribution is stronger than infinite strings remains unchanged.}
Loops can emit GWs with frequency $\omega \gtrsim (\alpha t)^{-1}$,
where $\alpha$ is the parameter which represents the typical loop size
normalized by the horizon scale.  Therefor, unless $\alpha$ is much less
than $1$, GWs from loops dominate in the wide range of frequency bands
detected by GW detection experiments, e.g. pulsar timing or ground
based or space-borne GW detectors.
 
However, there are several reasons to consider GWs of infinite strings.
First, recent papers suggest that $\alpha$  
is much less than previously thought~\cite{Siemens:2002dj,Polchinski:2006ee},
and hence loops cannot emit low frequency GWs which can be detected 
by pulsar timing experiments.
If this is true, 
GWs from infinite strings can give dominant contribution to the GW
background at those low frequencies, as we will show in this paper. 
Second, infinite strings can emit GWs with wavelength of horizon scale
because the characteristic scale of infinite strings is as long as the size 
of the horizon. This may be detected by future and on-going CMB surveys.
Loops cannot emit GWs with such long wavelength.\footnote{%%
Other recent simulations imply $\alpha \sim 0.1$, which is much bigger 
than $G\mu$. In this case loops can emit GWs whose wavelength is 
comparable to or somewhat shorter than the horizon scale~\cite{Ringeval:2005kr}.}%
Thus, we are lead to consider GWs from infinite strings.

%There exist a couple of structures on infinite strings which are 
%responsible for GW emission.  One is the small scale structure of
%infinite strings.  The GW emission from the small scale structure was 
%originally discussed in order to derive the size of the
%smallest structure on infinite 
%string~\cite{Hindmarsh:1990xi,Siemens:2001dx,Siemens:2002dj}.
%\rem{(Why we do not consider this case?)}
%, and  
%few literatures considered the detectability of these GWs as a stochastic
%background. 
%Another structure which is considered as more important source of GWs is a 
%kink.
A structure on infinite strings which is responsible for GW emission is a kink.\footnote{%%.
Generically, infinite strings do not have any cusp since its existence depends
on the boundary condition of the string, i.e., the periodic condition. }%
Kinks are produced when infinite strings reconnect and kinks on the 
infinite strings can produce GW bursts. The kinks are sharp when they 
first appear, but are gradually smoothened as the Universe expands. 
Moreover, when an infinite string self-intercommutes 
and produces a loop, some kinks immigrate from the infinite string to the loop, 
and hence the number of kinks on the infinite string decreases. 
Recently, Copeland and Kibble derived the distribution function of 
kinks on an infinite string  taking into account these effects~\cite{Copeland:2009dk}.
However, they did not study the GWs from kinks. 

Therefore, in this paper,  we investigate GWs from
kinks using the kink distribution obtained in~\cite{Copeland:2009dk}
and discuss the detectability of them by future GW experiments, 
especially by pulsar timing, such as Square Kilometer Array (SKA)
and space-based detectors such as DECIGO and BBO.

This paper is organized as follows. 
In section 2, we briefly review a part of the basis of cosmic strings. 
In section 3, we derive the distribution function of kinks on infinite strings according to
~\cite{Copeland:2009dk}. 
In section 4, we derive the formula of the energy radiated from kinks per unit time.
In section 5, we calculate the spectrum of stochastic GW background originating from kinks, using formula derived in section 4.
Section 6 is devoted to summary and discussion.

%%%%%%%%%%%%%%%%%%%%%%%%%%%%%%%%%%%%%%%%%
  \section{Dynamics of cosmic strings}
%%%%%%%%%%%%%%%%%%%%%%%%%%%%%%%%%%%%%%%%%

The dynamics of a cosmic string, whose width can be neglected, is described
by the Nambu-Goto action, 
\begin{equation}
  S=-\mu \int d^2 \zeta \sqrt{-\det(\gamma_{ab})}. \label{NGaction}
\end{equation}
where $\zeta^a \ (a=0,1)$ are coordinates on the world sheet of the cosmic 
string, $\gamma_{ab} = g_{\mu \nu} x^{\mu}_{, a} x^{\nu}_{,b}$ 
($x^{\mu}_{,a}=\frac{\partial x^{\mu}}{\partial \zeta^a}$ )  is the induced 
metric on the world sheet, and $\mu$ is the tension of the string. 
The energy-momentum tensor is 
\begin{equation}
   T^{\mu \nu}(x)=\mu \int d^2\zeta \sqrt{-\det(\gamma_{ab})} 
   \gamma^{ab}x^{\mu}_{,a}x^{\nu}_{,b} \delta^{4}(x-X(\zeta)), 	
   \label{EMtensor}
\end{equation}
where $X=X(\zeta)$ is embedding of the world sheet on the background 
metric. 
If the background space-time is Minkowski one, we can select the coordinate 
system $(\zeta^0,\zeta^1)=(\tau, \sigma)$ which satisfies the gauge conditions 
\begin{equation}
   \tau = t \ (\mathrm{physical \ time}), 
   ~~~~~~\ x_{,\tau} \cdot x_{,\sigma}=0, \ 
   ~~~~~~x^2_{,\tau} + x^2_{,\sigma} =0.
\end{equation}
The time scale of a GW burst is much shorter than the Hubble expansion, 
hence we consider an individual burst event on the Minkowskian background.
The general solution of the equation of motion derived from the action
(\ref{NGaction}) is 
  \begin{equation}
  	x^{\mu} = \frac{1}{2} (a^{\mu}(u) + b^{\mu}(v)), \ 
	\mathbf{a}^{\prime 2}(u) = \mathbf{b}^{\prime 2}(v)=1
  \end{equation}
where $u=\sigma+t, v=\sigma-t$. We call $a(u)$ $(b(v))$ the 
left (right)-moving mode. 
Then, Eq.~(\ref{EMtensor}) can be rewritten in terms of $a(u)$ and $b(v)$,
\begin{equation}
	T^{\mu \nu}(k) =  \frac{\mu}{4}(I^{\mu}_+(k)I^{\nu}_-(k) 
	+ I^{\mu}_-(k)I^{\nu}_+(k)), 
	\label{EMtensor2}
\end{equation}
\begin{equation}
	I_+^{\mu}(k)=\int du a^{\prime\mu}(u) e^{ik\cdot a(u)/2}, \ 
	I_-^{\mu}(k) = \int dv b^{\prime\mu}(v) e^{ik\cdot b(v)/2}, \label{I}
\end{equation} 
where $T^{\mu \nu}(k)$ is the Fourier transform of the $T^{\mu \nu}(x)$, i.e.  
$T^{\mu \nu}(k) = \int d^4x T^{\mu \nu}(x)e^{ik\cdot x}$.

%%%%%%%%%%%%%%%%%%%%%%%%%%%%%%%%%%%%%%%%%
\section{Distribution function of kinks}
%%%%%%%%%%%%%%%%%%%%%%%%%%%%%%%%%%%%%%%%%

Kinks can be defined as discontinuities of $\mathbf{a}^{\prime}$ or 
$\mathbf{b}^{\prime}$. 
They are produced when two infinite strings collide and reconnect 
because $\mathbf{a}^{\prime}$ and $\mathbf{b}^{\prime}$ on the new infinite 
string are created by connecting $\mathbf{a}^{\prime}$s or 
$\mathbf{b}^{\prime}$s on two different strings. 
Let us suppose that $\aprime$ jumps from $\aprime_-$ to $\aprime_+$  
at a kink. Then the  ``sharpness'' of a kink is defined by
 \be
	 \psi = \frac{1}{2}(1-\aprime_+ \cdot \aprime_-).
 \ee
Thus the norm of the difference between $\aprime_-$ and $\aprime_+$ is 
$|\Delta \aprime | = 2\sqrt{\psi}$. 
The production rate of kinks is given by~\cite{Copeland:2009dk}
  \be
  	\dot{N}_{\rm production} = \frac{\bar{\Delta}V}{\gamma^4t^4} g(\psi)	,
  \ee
where $N(t,\psi)d\psi$ denotes the number of kinks 
with sharpness between $\psi$ and $\psi+d\psi$ in the volume $V$, 
$\bar{\Delta}$ and $\gamma$ are constants related to string networks,
whose values are~\cite{Copeland:2009dk},
  \be
  	\bar{\Delta}_r \simeq 0.20, \ \bar{\Delta}_m \simeq 0.21, \  
	\gamma_r\simeq0.3, \ \gamma_m\simeq0.55,
  \ee
Here the subscript $r$($m$) denotes the value in the 
radiation(matter)-dominated era.
\be
g(\psi)=\frac{35}{256}\sqrt{\psi}(15-6\psi -\psi^2) \label{g}
\ee
 and we set $g(\psi)=0$ for $\psi<0, 1<\psi$.
The correlation length of the cosmic strings $\xi$ is given by 
$\xi \simeq \gamma t$.
  
Produced kinks are blunted by the expansion of the Universe. 
The blunting rate of the kink with the sharpness $\psi$ is given 
by~\cite{Copeland:2009dk}
  \be
	  \frac{\dot{\psi}}{\psi}\bigg|_{\rm stretch} = -2\zeta t^{-1},
  \ee
where $\zeta$ is a constant which, in the radiation(matter)-dominated era, 
is given by $\zeta_r \simeq 0.09~  (\zeta_m \simeq 0.2)$.

On the other hand, the number of kinks on an infinite string decreases 
when it self-intercommutes since some kinks are taken away by the loop 
produced. The decrease rate of kinks due to this effect is given 
by~\cite{Copeland:2009dk}
 \be
	   \frac{\dot{N}}{N}\bigg|_{\rm to~loop}=-\frac{\eta}{\gamma t},
 \ee
where $\eta$ is constant which, in the radiation(matter)-dominated era,
is given by $\eta_r \simeq 0.18 ~(\eta_m \simeq 0.1$).

Taking into account these effects, the evolution of the kink number $N$ 
obeys the following equation,
 \be
    \dot N = \frac{\bar\Delta V}{\gamma^4 t^4}g(\psi) + 
 	\frac{2\zeta}{t}\frac{\partial}{\partial\psi}(\psi N) - 
 	\frac{\eta}{\gamma t}N. \label{Neq}
 \ee
 %%
 
  %%%%%%%%%%%%%%%%  table %%%%%%%%%%%%%%%%%%%%%%
\begin{table}[t]
  \begin{center}
    \begin{tabular}{ | c | c | c | c | }
      \hline 
       ~ & definition  & radiation dom. & matter dom. 
      \\    \hline 
      $\gamma$ & -                            & 0.3           & 0.55 \\
      $\zeta$       & -                            & 0.09         & 0.2 \\
      $\beta$       & -                            & 1.1           & 1.2 \\
      $A$               & $2\zeta-\beta$ & $-0.92$  & $-0.8$ \\
      $B$               & $4\zeta-\beta$ & $-0.74$  & $-0.4$ \\
      $C$               & $(\beta-8\zeta)/[3(\beta-2\zeta)]$ & 0.14  & $-0.17$ \\
      \hline 
    \end{tabular}
    \caption{ 
          Various constants appearing in the calculation of kink distribution and
          the spectrum of GWs from kinks.
    }
    \label{table:constants}
  \end{center}
\end{table}
%%%%%%%%%%%%%%%%%%%%%%%%%%%%%%%%%%%%%%%%%%%%%% 

We have to solve this equation under an appropriate initial condition. 
The initial condition to be imposed depends on how strings emerge. 
We consider two typical scenarios.
As a first scenario, let us suppose that
cosmic strings form when spontaneous symmetry breaking (SSB) occurs
in the radiation dominated Universe.
After the formation, cosmic strings interact with particles in thermal bath,
which acts as friction on the string.
At first this friction effect overcomes the Hubble expansion.
(This epoch is called friction-dominated era.)
The important observation is that the friction effect smoothens strings 
and washes out the small scale structure on them~\cite{Garriga:1993gj}.
The temperature at which friction domination ends is given by~\cite{Vilenkin:1991zk}
 \be
	 T_c \sim G\mu M_{\rm pl}, \label{Tefd}
 \ee
where $G$ is the Newton constant and $M_{\rm pl}$ is the Planck scale.
If the temperature at the string formation is lower than this critical 
value, 
the friction effect can be neglected from the formation epoch to the present 
day, and kinks emerge at the same time as appearance of strings. 

As a second scenario, let us suppose that cosmic strings are generated 
at the end of inflation, like strings made by condensation of waterfall 
fields in supersymmetric hybrid 
inflation~\cite{Dvali:1994ms,Binetruy:1996xj,Jeannerot:1997is}, 
and cosmic superstrings left after annihilation of the D-brane and 
anti D-brane in brane inflation~\cite{Sarangi:2002yt,Dvali:2003zj}.
In this case, cosmic strings form in the inflaton-oscillation dominated era, 
which resembles the matter-dominated era.
Although strings feel frictions from dilute plasma existing before the 
reheating completes, the effect is negligible as long as the reheating 
temperature after inflation $T_{r}$ is lower than $\sim G\mu M_{\rm pl}$.
In this case, kinks begin to be formed right after the formation of strings.
Otherwise, the temperature at which kinks begin to appear is given by 
Eq.~(\ref{Tefd}).

In anyway, even if such kinks survive friction, they become extremely 
dense in the later period, so that they do not affect the observable 
part of the GW spectrum, as we will see later. 
On the other hand, if the reheating temperature is lower than $T_c$, 
the kinks from the first matter era definitely continue to exist, 
and if the reheating temperature is extremely low, GWs from these kinks 
can be observed.
Therefore, we concentrate on the situation where the reheating temperature 
is low enough so that the kinks generated in the first matter era survive 
without experiencing the friction domination. 
We denote the time when kinks starts to be formed by $t_*$. 
%For strings associated with SSB, $t_*$ may correspond to the end of the 
%friction domination, and for strings formed at the end of inflation, $t_*$ 
%may be just the end of inflation, although it depends on the thermal history after inflation.

Then we can get the solution, but its precise form is very complicated.
Here we assume that the matter-dominated epoch follows after inflation, and
the reheating completes at $t=t_r$ after which the radiation dominated era 
begins.
If we focus on only the dominant term and neglect $\mathcal{O}(1)$ numerical 
factor, we get
\be 
 \frac{dN}{d\psi}(t,\psi) \sim
	\begin{cases}
		\psi^{-\beta_m/2\zeta_m} t^{-1}
		& {\rm for~~}\psi> \left( \frac{t_*}{t} \right)^{2\zeta_m} \\
		\left( \frac{t}{t_*} \right)^{\beta_m+\zeta_m}\psi^{1/2} t^{-1} 
		& {\rm for~~}\psi< \left( \frac{t_*}{t} \right)^{2\zeta_m}
	\end{cases} 
	\label{bunpueasymat1}
\ee
in the first matter era,
\be
 \frac{dN}{d\psi}(t,\psi) \sim
	 \begin{cases}
		\psi^{-\beta_r/2\zeta_r} t^{-1}
		& {\rm for~~}\psi> \psi^{\rm (RD)}_1 (t) \\
		\left( \frac{t}{t_r} \right)^{\beta_r-\beta_m\zeta_r/\zeta_m}
		\psi^{-\beta_m/2\zeta_m} t^{-1} 
		& {\rm for~~}\psi^{\rm (RD)}_2 (t) 
		<\psi< \psi^{\rm (RD)}_1 (t) \\
		\psi^{1/2} \left( \frac{t}{t_r}\right)^{\beta_r+\zeta_r}
		 \left( \frac{t_r}{t_*} \right)^{\beta_m+\zeta_m}t^{-1} 
		 &  {\rm for~~}\psi < \psi^{\rm (RD)}_2 (t)
	\end{cases} 
	\label{bunpueasyrad}
\ee
in the radiation era where
\begin{eqnarray}
	&\psi^{\rm (RD)}_1 (t)&=  \left( \frac{t_r}{t} \right)^{2\zeta_r},\\
	&\psi^{\rm (RD)}_2 (t)&= 
	\left(\frac{t_r}{t}\right)^{2\zeta_r} \left(\frac{t_*}{t_r}\right)^{2\zeta_m}, 
\end{eqnarray}
and
\begin{align}
 \frac{dN}{d\psi}(t,\psi) \sim 
 &
	 \begin{cases}
		\psi^{-\beta_m/2\zeta_m} t^{-1}
		& {\rm for~~}\psi >  \psi^{\rm (MD)}_1(t)\\
		\left( \frac{t}{t_{eq}} \right)^{\beta_m-\beta_r\zeta_m/\zeta_r}
		\psi^{-\beta_r/2\zeta_r} t^{-1} 
		&{\rm for~~}  \psi^{\rm (MD)}_2(t)
		<\psi<  \psi^{\rm (MD)}_1(t) \\
		\left( \frac{t_{eq}}{t_r} \right)^{\beta_r-\beta_m\zeta_r/\zeta_m}
		\psi^{-\beta_m/2\zeta_m} t^{-1} 
		&{\rm for~~} \psi^{\rm (MD)}_3(t) <\psi < \psi^{\rm (MD)}_2(t) \\
		\psi^{1/2} \left(\frac{t}{t_{eq}}\right)^{\beta_m+\zeta_m}
		\left( \frac{t_{eq}}{t_r}\right)^{\beta_r+\zeta_r} 
		\left( \frac{t_r}{t_*} \right)^{\beta_m+\zeta_m}t^{-1} 
		 &{\rm for~~}\psi < \psi^{\rm (MD)}_3(t)
	\end{cases} \label{bunpueasymat2}
\end{align}
in the second matter era,
where
\begin{eqnarray}
	&\psi^{\rm (MD)}_1 (t)&= \left( \frac{t_{eq}}{t} \right)^{2\zeta_m},
	\label{psiMD1} \\
	&\psi^{\rm (MD)}_2 (t)&= 
	\left(\frac{t_{eq}}{t}\right)^{2\zeta_m} 
	\left(\frac{t_r}{t_{eq}}\right)^{2\zeta_r}, \\
	&\psi^{\rm (MD)}_3 (t)&=  
	\left(\frac{t_{eq}}{t}\right)^{2\zeta_m}\left(\frac{t_r}{t_{eq}}\right)^{2\zeta_r} 
		\left(\frac{t_*}{t_r}\right)^{2\zeta_m}.
	\label{psiMD3}
\end{eqnarray}
Here $\beta$ is the constant related to the string network ($\beta_r 
\simeq 1.1, \ \beta_m \simeq 1.2$), and $t_{eq}$ denotes the matter-radiation equality epoch.
We have converted $N$, distribution in the volume $V$, to 
$dN/d\psi= N(t,\psi)/(V/\xi^2)$, which is the distribution per unit length.
The derivation of the above expression of $dN/d\psi$ is described in Appendix~\ref{sec:app0}.

%%%%%%%%%%%%%%%%%%%%%%%%%%%%%
\begin{figure}[t]
\begin{center}
\includegraphics[width=1\linewidth]{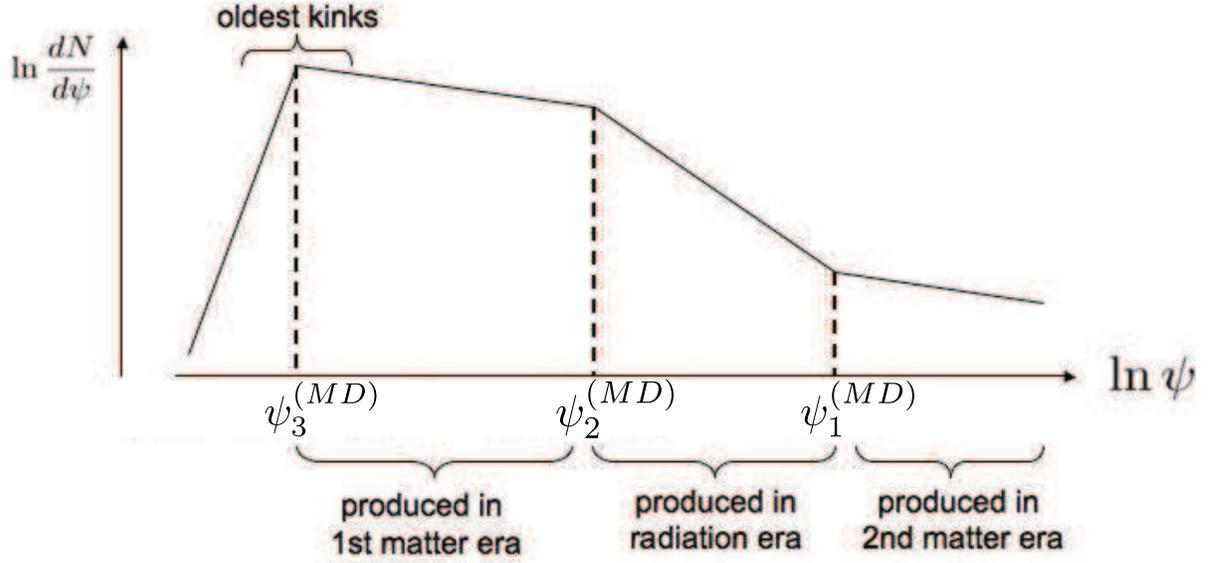}
\caption{The distribution function of kinks on infinite strings produced 
at the end of the inflation in the second matter era. 
$\psi_1^{\rm (MD)}$, $\psi_2^{\rm (MD)}$, $\psi_3^{\rm (MD)}$ are given by 
Eqs.~(\ref{psiMD1})-(\ref{psiMD3}). }
\label{fig:kinkdist}
\end{center}
\end{figure}
%%%%%%%%%%%%%%%%%%%%%%%%%%%%%

Let us consider the physical meaning of this distribution function. 
It is not difficult to consider how the number of kinks in the horizon
changes as time goes on. 
When a kink is born, its sharpness ranges from 0 to 1, 
but the typical value is $\mathcal O(0.1)$. 
Therefore, the kink distribution has a peak at  $\psi\sim 
\mathcal O(0.1)$ at the very early stage. 
Then kinks are made blunt by the cosmic expansion.  
Thus newly produced kinks are sharp (large $\psi$) and old ones are blunt 
(small $\psi$).
Therefore, the peak value of $\psi$ is much smaller than 0.1 at the 
late stage.
This peak consists of the oldest kinks. 
Fig.~\ref{fig:kinkdist} roughly sketches the shape of the distribution 
function in the second matter era.

It should be noted that the above distribution is derived 
without considering gravitational backreaction. 
Since the most abundant kinks are extremely blunt, 
they might be influenced by the gravitational backreaction and disappear. 
However, it is difficult and beyond the scope of this paper to take backreaction 
into account.
We will comment on this issue in Appendix~\ref{sec:appB}.

%%%%%%%%%%%%%%%%%%%%%%%%%%%%%%%%%%%%%%%%%%%%%%
\section{THE SPECTRUM OF GRAVITATIONAL WAVES FROM KINKS}
%%%%%%%%%%%%%%%%%%%%%%%%%%%%%%%%%%%%%%%%%%%%%%

Given the energy-momentum tensor of the source, 
one can calculate the energy of the GW in the direction of $\hat{k}$
with frequency $\omega$ as~\cite{CandG}
\begin{eqnarray}
	\frac{dE}{d\Omega}(\hat{k}) & = 
	& 2G\int^{\infty}_{0}d\omega \omega^2
	 \left( T^{\mu \nu *}(k)T_{\mu \nu}(k) 
	 - \frac{1}{2}|T^{\mu}_{\mu}(k)|^2 \right) \nonumber \\
	& = &2G \Lambda_{ij.lm}(\hat{k})\int^{\infty}_0 d\omega \omega^2 
	T^{ij*}(k)T^{lm}(k)
\end{eqnarray}
\be
	\Lambda_{ij,lm}(\hat{k}) \equiv 
	\delta_{il} \delta_{jm} - 2\kh_j\kh_m\delta_{il} +
	\frac{1}{2}\kh_i\kh_j\kh_l\kh_m-\frac{1}{2}\delta_{ij}\delta_{lm} 
	+ \frac{1}{2}\delta_{ij}\kh_l\kh_m 
	+\frac{1}{2}\delta_{lm}\kh_i\kh_j
\ee

Thus by substituting the energy-momentum tensor (\ref{EMtensor2}), 
we find the energy of the GW after computing integral $I_{\pm}$
in Eq.~(\ref{I}). 
It is given by
\begin{align}
	\frac{dE}{d\omega d\Omega} (k)= \frac{G\mu^2\omega^2}{4} 
	\bigg(& |\Ip|^2|\Imi|^2 + |\Ip \cdot \Imi|^2 - |\Ip\cdot\Imi^*|^2 
	\nonumber \\ 
	&-|\Ip|^2|\kh \cdot \Imi|^2-|\Imi|^2|\kh\cdot\Ip|^2   
	+ |\kh\cdot\Ip|^2|\kh\cdot\Imi|^2 \nonumber \\
	&- (\Ip\cdot\Imi^*)(\kh\cdot\Ip^*)(\kh\cdot\Imi) 
	- (\Ip^*\cdot\Imi)(\kh\cdot\Ip)(\kh\cdot\Imi^*) 
	\nonumber \\
	&+ (\Ip\cdot\Imi)(\kh\cdot\Ip^*)(\kh\cdot\Imi^*)
	+(\Ip^*\cdot\Imi^*)(\kh\cdot\Ip)(\kh\cdot\Imi) \bigg) 
	\label{Egeneral}
\end{align}
where $\vec{I}_{\pm} \equiv (I^1_{\pm}(k),I^2_{\pm}(k),I^3_{\pm}(k))$.

The method to calculate $I_{\pm}$ is described 
in~\cite{Damour:2001bk,Binetruy:2009vt}. 
In the limit $\omega \rightarrow \infty$, $I_{\pm}$ exponentially reduces to $0$, 
unless at least one of the following conditions on the integrand is met.
One is the existence of discontinuities of $a^{\prime i}(u)$ (or $b^{\prime i}(v)$). 
The contribution of a discontinuity of $\aprime$ to $I_+(k)$ is
\be
	I^i_+(k) \simeq -\frac{2}{i\omega}
	\left(\frac{a^{\prime i}_+}{1-\kh\cdot\aprime_+} 
	- \frac{a^{\prime i}_-}{1-\kh\cdot\aprime_-} \right)
	e^{i\omega(u_*-\kh\cdot\mathbf{a}_*)/2} 
	\label{Idisc}
\ee
where $u_*$ is the position of the discontinuity, $\mathbf{a}_*=\mathbf{a}(u_*)$
and we assume $\aprime$ jumps from $\aprime_-$ to $\aprime_+$ at $u=u_*$. 
The region of length $\sim\omega^{-1}$ around $u=u_*$ contributes to this value. 
We find $I^i_+(k) \sim \psi/\omega$, where $\psi$ denotes 
sharpness of the kink.
The other condition is the existence of stationary points of the phase of the 
integrand, i.e. 
$\omega (u-\kh\cdot\mathbf{a}(u))/2$ (or $\omega (-v-\kh\cdot\mathbf{b}(v))/2$). 
This condition is expressed as
\be
	1-\kh\cdot\aprime(u_s)=0 \ (\mathrm{or} \ -1-\kh\cdot\bprime(v_s)=0) \label{sta-con}
\ee
at the point $u=u_s$($v=v_s$). The contribution of the stationary point of the phase to $I_+(k)$ is
\be
	I^i_+(k) \simeq \frac{1}{\omega^{2/3}} a^{\prime \prime}_i(u_s)
	e^{i\omega(u_s-\kh\cdot\mathbf{a}(u_s))/2} 
	\left( \frac{12}{|\kh\cdot\mathbf{a}^{\prime \prime \prime}(u_s)|}\right)^{2/3}
	\frac{i}{\sqrt{3}}\Gamma(2/3). \label{Ista}
\ee
For any value of $\aprime \ (\bprime)$, there is one direction $\kh$ that 
satisfies (\ref{sta-con}), i.e. $\kh = \aprime \ (-\bprime)$. 
Therefore, every point of $u$ can contribute to $I^i_+(k)$ for one direction $\kh$.

One can find the energy of the GW burst from ONE kink by picking up the contribution 
from the discontinuity for one of $I_{\pm}$ (say, $I_+$) and 
from the stationary point for the other (say, $I_-$). 
The energy emitted when the kink is located at the world sheet coordinate 
$(u,v)=(u_*,v_s)$ is evaluated by substituting (\ref{Idisc}) into $I_+$ and 
(\ref{Ista}) into $I_-$ in (\ref {Egeneral}). 
Then we obtain
\begin{align}
	\frac{dE}{d\Omega d\omega}(\omega,-\omega \bprime_s)=&
	\frac{\Gamma ^2\left(\frac{2}{3}\right)}{3}G\mu^2 \omega^{-\frac{4}{3}} 
	\left(\frac{12}{|\bprime_s\cdot\mathbf{b}^{\prime \prime \prime}_s|} 
	\right)^{\frac{4}{3}} \nonumber \\
	&\times
	\left(\frac{1}{(1+\bprime_s\cdot\mathbf{a}^{\prime}_{+} )^2}-
	\frac{2 \mathbf{a}^{\prime}_+ \cdot\mathbf{a}^{\prime}_-}
	{(1+\bprime_s\cdot\mathbf{a}^{\prime}_{+} )	
	(1+\bprime_s\cdot\mathbf{a}^{\prime}_{-} )} 
	+\frac{1}{(1+\bprime_s\cdot\mathbf{a}^{\prime}_{-} )^2} 
	\right)\mathbf{b}^{\prime \prime 2}_s,
\end{align}
where the subscript $s$ represents the value at $v=v_s$.
The energy is radiated at every moment toward the direction of $-\bprime$ from 
the kink.
This is the formula which represents energy emitted in a short period in a small 
solid angle. 
The total energy emitted in the short period $\Delta t$ is found by multiplying 
$\Delta \Omega$, 
which is the solid angle that the GW sweeps in this short period. 
The extent of the radiation has the solid angle 
$\sim ((\omega / |\mathbf{b}^{\prime \prime}_s|)^{-1/3})^2$~\cite{Damour:2001bk}. 
The variation of the direction of the GW is roughly estimated by 
$|\Delta \bprime| \sim |\mathbf{b}^{\prime \prime} \Delta t|$. 
As a result, we obtain
\be
	\Delta \Omega \sim (\omega  / |\mathbf{b}^{\prime \prime}_s|)^{-1/3}
	|\mathbf{b}^{\prime \prime}_s|\Delta t
\ee
and the energy emitted per unit time is 
\be
\begin{split}
	 \frac{dP}{d\omega}  \sim
	&\frac{\Gamma ^2(2/3)}{3}G\mu^2 \omega^{-5/3}
	 \left(\frac{12}{|\bprime_s\cdot\mathbf{b}^{\prime \prime \prime}_s|} 
	 \right)^{4/3} \\
	 &\times
	 \left(\frac{1}{(1+\bprime_s\cdot\mathbf{a}^{\prime}_{+} )^2}-
	 \frac{2 \mathbf{a}^{\prime}_+ \cdot\mathbf{a}^{\prime}_-}
	 {(1+\bprime_s\cdot\mathbf{a}^{\prime}_{+} )
	 (1+\bprime_s\cdot\mathbf{a}^{\prime}_{-} )} 
	 +\frac{1}{(1+\bprime_s\cdot\mathbf{a}^{\prime}_{-} )^2} 
	 \right)|\mathbf{b}^{\prime \prime }_s|^{10/3}, 
	 \label{dPdomega}
\end{split}
\ee
The terms in the large parenthesis in the 2nd line of Eq.~(\ref{dPdomega}) 
can be estimated by taking average 
over the angle between the left- and right-moving mode as
$\langle \frac{1}{(1+\bprime_s \cdot \aprime_{\pm})^2} \rangle \sim  
\langle \frac{1}{(1+\bprime_s \cdot \aprime_+)(1+\bprime_s\cdot\aprime_-)} 
\rangle \sim  1$.
The magnitudes of $\mathbf{b}^{\prime \prime}_s$ and $\mathbf{b}^{\prime 
\prime \prime}_s$ should be $\xi^{-1} (\sim t^{-1})$ and $\xi^{-2} (\sim t^{-2})$.\footnote{
	Although $\mathbf{b}^{\prime}$  has small and dense discontinuities which represent kinks,  
$v_s$ is 
generally not  a kink point. So the rough shape of  $\mathbf{b}_s^{\prime}$ is determined by 
the global 
appearance of the string network and hence the length scale of its 
variation is roughly 
the curvature radius of the network.
}

After making these substitutions and neglecting $\mathcal{O}(1)$ numerical 
factors, we find
\be
	\frac{dP}{d\omega} \bigg|_{\mathrm{one \ kink}} 
	\sim 10  G \mu^2 \psi \omega^{-5/3} t^{-2/3}.
\ee

So far we have evaluated the GW spectrum from one kink.
However, there exist many kinks on an infinite string and the 
final observable GW spectrum is made from sum of contribution from these kinks. 
Therefore, $I_+$ picks contributions of many kinks. Formally, 
\be
	I^i_+(k) = \sum_m I^i_{+,m}(k),
\ee
where an integer $m$ labels each kink. 
Thus Eq.~(\ref{Egeneral}) has cross terms of the contributions from 
different kinks, e.g., $\vec{I}_{+,m} \cdot \vec{I}^*_{+,n}$. 
However, such cross terms must vanish since a GW burst from a kink is a local 
phenomenon which relates only the region around the kink. 
In fact, the structure of the string around a specific point arises as a result 
of nonlinear evolution of the string network, and hence the values of 
$\mathbf{b}_s^{\prime \prime}$, $\mathbf{b}_s^{\prime \prime \prime}$ and so on, 
are stochastic.
We show that ensemble averages of cross terms vanish as expected, i.e. 
$\langle I^i_{+,m}I^{j*}_{+,n}\rangle=0$ in Appendix~\ref{sec:appA},
where it is also shown that the kinks that dominantly contribute to 
the power of GWs with frequencies $\sim \omega$ are ones which satisfy
\be
	\left(\psi \frac{dN}{d\psi} \right)^{-1} \sim \omega^{-1}. 
	\label{kinkcon}
\ee
In other words, if the interval of kinks with sharpness $\sim \psi$ is similar 
to the period of the GW under consideration, these kinks make dominant 
contribution to the GW. 
In the first matter era, using Eq.~(\ref{bunpueasymat1}), 
Eq.~(\ref{kinkcon}) 
simplifies to\footnote{%%
Strictly speaking, Eq.~(\ref{kinkcon}) has two solution for $\psi$, 
but it is sufficient to take larger one.}%%
\be
	\psi \sim (\omega t)^{2\zeta_m/A_m}  
\ee
for $\omega < (t_*/t)^{A_m}t^{-1}$. 
(Here, we set $A \equiv 2\zeta-\beta$. $A_r \simeq -0.92, \ A_m\simeq -0.8$.) 
For $\omega > (t_*/t)^{A_m}t^{-1}$, 
$\left(\psi \frac{dN}{d\psi} \right)^{-1} > \omega^{-1}$ is satisfied by 
an arbitrary value of $\psi$. 
In Appendix~\ref{sec:appA} it is shown that for  $\omega > (t_*/t)^{A_m}t^{-1}$ 
the main contribution to $\omega \frac{dP}{d\omega}$ comes from the kinks 
corresponding to  the peak of the distribution, i.e. $\psi \sim (t_*/t)^{2\zeta_m}$. 
If we denote  the value of sharpness of kinks which make dominant contribution 
to $\omega \frac{dP}{d\omega}$ as $\psi_{\rm max}(\omega,t)$, it is given by
\be
	\psi_{\rm max}(\omega,t) \sim
	\bc
	(\omega t)^{2\zeta_m/A_m}   
	& {\rm for~~}\omega < (t_*/t)^{A_m}t^{-1} \\
	\left(\frac{t_*}{t}\right)^{2\zeta_m}  
	& {\rm for~~}\omega > (t_*/t)^{A_m}t^{-1}
	\ec \label{psimaxmat1}
\ee
in the first matter era. In the radiation era, $\psi_{\rm max}(\omega,t)$ is 
found in a similar way as
\be
\psi_{\rm max}(\omega,t) \sim
\bc
	(\omega t)^{2\zeta_r/A_r} 
	& {\rm for~~}\omega<\omega^{\rm (RD)}_1 (t)\\
	\left( \frac{t_r}{t} \right)^{2D/A_m}(\omega t)^{2\zeta_m/A_m} 
	& {\rm for~~}\omega^{\rm (RD)}_1 (t) < \omega < \omega^{\rm (RD)}_2 (t) \\
	 \left(\frac{t_r}{t}\right)^{2\zeta_r} \left(\frac{t_*}{t_r}\right)^{2\zeta_m} 
	 & {\rm for~~}\omega >\omega^{\rm (RD)}_2(t),
\ec. \label{psimaxrad}
\ee
where
\begin{eqnarray}
	&\omega^{\rm (RD)}_1 (t)&= \left( \frac{t_r}{t}\right)^{A_r}t^{-1},\\
	&\omega^{\rm (RD)}_2 (t)&= \left(\frac{t_r}{t} \right)^{A_r} \left(\frac{t_*}{t_r} \right)^{A_m}t^{-1}.
\end{eqnarray}
In the second matter era, $\psi_{\rm max}$ is estimated as
\be
	\psi_{\rm max}(\omega,t) \sim
	\bc
 	(\omega t)^{2\zeta_m/A_m} 
	&{\rm for~~}\omega<\omega^{\rm (MD)}_1(t) \\
	 \left( \frac{t_{eq}}{t} \right)^{-2D/A_r}(\omega t)^{2 \zeta_r/A_r} 
	 &{\rm for~~}\omega^{\rm (MD)}_1(t) < \omega < \omega^{\rm (MD)}_2(t) \\
	 \left( \frac{t_r}{t_{eq}} \right)^{2D/A_m}(\omega t)^{2\zeta_m/A_m} 
	&{\rm for~~}\omega^{\rm (MD)}_2(t)< \omega < \omega^{\rm (MD)}_3(t) \\
  	\left(\frac{t_{eq}}{t}\right)^{2\zeta_m} 
	\left(\frac{t_r}{t_{eq}}\right)^{2\zeta_r}\left(\frac{t_*}{t_r}\right)^{2\zeta_m}  
	&{\rm for~~}\omega > \omega^{\rm (MD)}_3(t),
\ec. \label{psimaxmat2}
\ee
where
\begin{eqnarray}
	&\omega^{\rm (MD)}_1 (t)&= \left( \frac{t_{eq}}{t}\right)^{A_m}t^{-1},\\
	&\omega^{\rm (MD)}_2 (t)&= 
	\left(\frac{t_{eq}}{t} \right)^{A_m} \left(\frac{t_r}{t_{eq}} \right)^{A_r}t^{-1}, \\
	& \omega^{\rm (MD)}_3 (t)&= 
	\left(\frac{t_{eq}}{t} \right)^{A_m} \left(\frac{t_r}{t_{eq}} \right)^{A_r}
	\left(\frac{t_*}{t_r}\right)^{A_m}t^{-1}.
\end{eqnarray}
Here, we set $D\equiv \beta_r\zeta_m-\beta_m\zeta_r \simeq 0.11$.

As a result, assuming that the powers of GW from different kinks are  roughly 
same as far as their sharpnesses are in the same order, (in other words, 
assuming that the quantities concerned with kinks, such as 
$\mathbf{b}^{\prime \prime}_s$, except their sharpness, are roughly same) 
we can estimate the total power of GWs with frequencies $\sim \omega$ from 
all of the kinks in a horizon as, 
 \be
 	\omega \frac{dP}{d\omega} \bigg|_{\rm tot} \sim 
 	\omega \frac{dP}{d\omega} \bigg|_{\mathrm{one \ kink}} (\omega,\psi_{\rm max}(\omega,t))
 	\times \psi \frac{dN}{d\psi} \bigg|_{\psi = \psi_{\rm max}(\omega,t)} \times t. 
	\label{omedPdome}
\ee
The first factor denotes the power of GWs from one kink. 
The second factor denotes the number of kinks which satisfy 
$\psi \sim \psi_{\rm max}(\omega,t)$ per unit length. 
The third factor is length of an infinite string in a horizon.
Then we finally obtain
 \be
 	\omega \frac{dP}{d\omega} \bigg|_{\rm tot} \sim
	 \bc
	 10G\mu^2(\omega t)^{C_m} 
	 & {\rm for~~}t^{-1}<\omega < (t_*/t)^{A_m}t^{-1}\\
 	10G\mu^2\left(\frac{t_*}{t} \right)^{B_m}(\omega t)^{-2/3} &  
	{\rm for~~}\omega > (t_*/t)^{A_m}t^{-1}
	\ec \label{powtotmat1}
\ee
in the first matter era, 
 \be
	  \omega \frac{dP}{d\omega} \bigg|_{\rm tot} \sim
	 \bc
 	10G\mu^2(\omega t)^{C_r} 
	&{\rm for~~}t^{-1}<\omega<\omega^{\rm (RD)}_1(t) \\
 	10G\mu^2 \left(\frac{t_r}{t} \right)^{2D/A_m}(\omega t)^{C_m}  
	& {\rm for~~} \omega^{\rm (RD)}_1(t) < 
	\omega < \omega^{\rm (RD)}_2(t) \\
 	10G \mu^2 \left( \frac{t_*}{t_r}\right)^{B_m} 
	\left( \frac{t_r}{t}\right)^{B_r}(\omega t)^{-2/3} 
	&{\rm for~~}\omega > \omega^{\rm (RD)}_2(t)
	\ec 
	\label{powtotrad}
\ee
in the radiation era, and
 \begin{align}
	 \omega \frac{dP}{d\omega} \bigg|_{\rm tot} \sim 
 	\bc
	10G\mu^2(\omega t)^{C_m} 
	&{\rm for~~}t^{-1}<\omega<\omega^{\rm (MD)}_1(t)  \\
	10G\mu^2 \left(\frac{t_{eq}}{t} \right)^{-2D/A_r}(\omega t)^{C_r} 
	&{\rm for~~} \omega^{\rm (MD)}_1(t)< \omega < \omega^{\rm (MD)}_2(t)\\
 	10G\mu^2 \left(\frac{t_{eq}}{t_r} \right)^{-2D/A_m}(\omega t)^{C_m} 
 	&{\rm for~~} \omega^{\rm (MD)}_2(t) < \omega 
	< \omega^{\rm (MD)}_3(t) \\
 	10G \mu^2 \left( \frac{t_*}{t_r}\right)^{B_m} 
	\left( \frac{t_r}{t_{eq}}\right)^{B_r}
	\left(\frac{t_{eq}}{t}\right)^{B_m}(\omega t)^{-2/3}  
	&{\rm for~~} \omega >\omega^{\rm (MD)}_3(t)
    \ec \label{powtotmat2}
\end{align}
in the second matter era, where 
$B\equiv 4\zeta-\beta~~(B_r\simeq -0.74, B_m\simeq -0.4)$ and 
$C\equiv (\beta-8\zeta)/3(\beta-2\zeta)~~(C_r\simeq 0.14,C_m\simeq-0.17$). 
A robust lower bound on the frequency of GWs is set to be $t^{-1}$ 
corresponding to the horizon scale.
%, since no causal processes can emit GWs of wavelength longer than this.

%%%%%%%%%%%%%%%%%%%%%%%%%%%%%%%%%%%%%%%%%%%%%%%%%%% 
\section{THE STOCHASTIC BACKGROUND OF GRAVITATIONAL WAVES FROM KINKS}
%%%%%%%%%%%%%%%%%%%%%%%%%%%%%%%%%%%%%%%%%%%%%%%%%%%

We have found Eqs.~(\ref{powtotmat1}),(\ref{powtotrad}) and 
(\ref{powtotmat2}) as the total energy radiated per unit time 
in a horizon from kinks on an infinite string.
Now we are in a position to calculate the density parameter of GWs defined by
\be
	\Omega_{\rm gw}(\omega) \equiv 
	\frac{\omega}{\rho_c} \frac{d\rho}{d\omega}(\omega),
\ee
where $\rho_c$ denotes the critical energy density of the present
Universe. 
Noting that the energy density of GWs decreases as $a^{-4}$
and the frequency redshifts as $a^{-1}$, we get
\be
	\Omega_{\rm gw}(\omega)  \sim 
	\frac{1}{\rho_c} \int^{t_0}_{t_*} dt \frac{1}{t^3} \left(\omega^{\prime}
	\frac{dP}{d\omega^{\prime}}\right) \bigg|_{\omega^{\prime} 
	= \omega \times a_0/a(t)} 
	\left(\frac{a(t)}{a_0}\right)^4, \label{forOmega}
\ee
where $a(t)$ is the scale factor and $a_0$ represents its present value. 
Using Eqs.~(\ref{powtotmat1}),(\ref{powtotrad}) and (\ref{powtotmat2}), 
we get the spectrum as
\begin{align}
   &\Omega_{\rm gw} (\omega)\sim \nonumber \\
   &
   \begin{cases}
	  60\pi(G\mu)^2(\omega t_0)^{C_m} 
	  \qquad \quad \ \quad \qquad \qquad 
	  &{\rm for~} t_0^{-1}<\omega< \omega_1 \\
	  60\pi(G\mu)^2\left(\frac{\Omega_r}{\Omega_m}\right)^{-3D/A_r}
	  (\omega t_0)^{C_r} 
 	   \qquad 
	   &{\rm for~} \omega_1 < \omega < \omega_2 \\
	  60\pi(G\mu)^2\left(\frac{\Omega_r}{\Omega_m}\right)^{-4D/A_m}
	  \left(\frac{T_0}{M_{\rm pl}}\right)^{4D/A_m}
	  \left(\frac{T_r}{M_{\rm pl}}\right)^{-4D/A_m}
	  (\omega t_0)^{C_m}
	  &{\rm for~} \omega_2 < \omega < \omega_3 \\
	  60\pi(G\mu)^2(\omega t_0)^{-2/3}
	  \left( \frac{\Omega_r}{\Omega_m}\right)^{3B_m/2-4B_r} 
	  \left(\frac{T_0}{M_{\rm pl}}\right)^{2B_r}
	  \left(\frac{T_r}{M_{\rm pl}}\right)^{2B_m-2B_r}
	  \left(\frac{H_*}{M_{\rm pl}}\right)^{-B_m}
	  &{\rm for~}\omega > \omega_3
   \end{cases}, \label{Omegagw}
\end{align}
where
\begin{eqnarray}
	&\omega_1 &= \omega^{\rm (MD)}_1(t_0) =
	\left( \frac{\Omega_r}{\Omega_m}\right)^{3A_m/2}t^{-1}_0,
	\label{omegaMD1} \\
	&\omega_2 &= \omega^{\rm (MD)}_2(t_0) =
	 \left( \frac{\Omega_r}{\Omega_m}\right)^{3A_m/2-2A_r}
	 \left(\frac{T_0}{M_{\rm pl}}\right)^{2A_r}
	 \left(\frac{T_r}{M_{\rm pl}}\right)^{-2A_r}t^{-1}_0, \\
	& \omega_3 &= \omega^{\rm (MD)}_3(t_0) =
	 \left( \frac{\Omega_r}{\Omega_m}\right)^{3A_m/2-2A_r}
	\left(\frac{T_0}{M_{\rm pl}}\right)^{2A_r}
	\left(\frac{T_r}{M_{\rm pl}}\right)^{2A_m-2A_r}
	\left(\frac{H_*}{M_{\rm pl}}\right)^{-A_m}t^{-1}_0
	\label{omegaMD3}
\end{eqnarray}
where $t_0$ and $T_0$ denote the present age and temperature of the Universe,  
$T_r$ is the reheating temperature
and $H_*$ is the Hubble parameter at the end of inflation.\footnote{
	Here we neglected the effect of cosmological constant.
	The GW spectrum will be slightly modified if the 
	cosmological constant is taken into account.
	Detailed estimation of this effect is beyond the scope of this paper
	since it needs a simulation of cosmic string network evolution in the
	cosmological constant dominated Universe.
}
This formula has complicated exponents. 
Using $\beta$ and $\zeta$ with values given above, 
$\Omega_m/\Omega_r \simeq 5.5\times10^3$ and 
$T_0/M_{\rm pl}\simeq 9.6\times 10^{-32}$, 
Eq.~(\ref{Omegagw}) simplifies to
 \begin{align}
	 \Omega_{\rm gw}(\omega) \sim 
 	&\bc
 	10^2(G\mu)^2(\omega t_0)^{-0.17}  
	 & {\rm for~~}t_0^{-1} < \omega < \omega_1\\
 	10(G\mu)^2(\omega t_0)^{0.14} 
	& {\rm for~~}\omega_1 < \omega 
	< \omega_2 \\
 	10^{18}(G\mu)^2\left(\frac{T_r}{M_{\rm pl}}\right)^{0.56}(\omega t_0)^{-0.17} 
	& {\rm for~~}\omega_2 <\omega
	< \omega_3 \\
 	10^{40}(G\mu)^2\left(\frac{T_r}{M_{\rm pl}}\right)^{0.68}
	\left(\frac{H_*}{M_{\rm pl}}\right)^{0.4} (\omega t_0)^{-2/3} 
	& {\rm for~~} \omega > \omega_3 ,
 \ec. 
 \end{align}
where
\begin{eqnarray}
	&\omega_1 & \sim 10^4t_0^{-1},\\
	&\omega_2 & \sim
	 10^{-3}\left(\frac{T_r}{T_0}\right)^{1.8}t_0^{-1}, \\
	& \omega_3 & \sim
	10^{54}\left(\frac{T_r}{M_{\rm pl}}\right)^{0.24}
	\left(\frac{H_*}{M_{\rm pl}}\right)^{0.8}t_0^{-1}
\end{eqnarray}
%%

%%%%%%%%%%%%%%%%%%%%%%%%%%%%%%%%%%%%%%
\begin{figure}[t]
\begin{center}
\includegraphics[width=18cm]{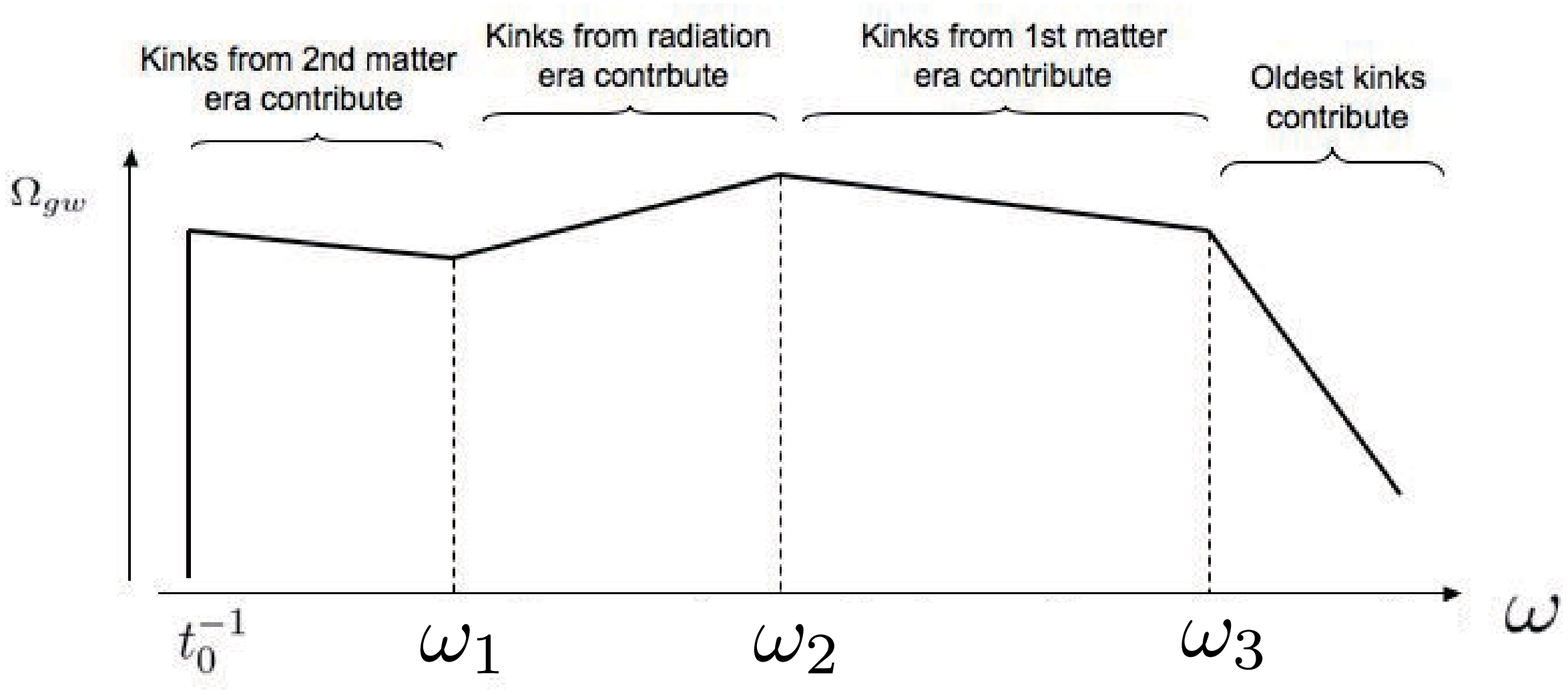}
\caption{A schematic picture of 
	gravitational wave spectrum from kinks (\ref{Omegagw}).
	$\omega_1$, $\omega_2$, $\omega_3$
	are given by Eqs.~(\ref{omegaMD1})-(\ref{omegaMD3}).
}
\label{fig:spectrum}
\end{center}
\end{figure}
%%%%%%%%%%%%%%%%%%%%%%%%%%%%%%%%%%%%%%

We find that the integral in Eq.~(\ref{forOmega}) is dominated by the 
contribution from the period near the present for the whole range 
of $\omega$. 
In other words, almost all of the present energy of GWs from 
the kinks on infinite strings comes from those radiated 
around the present epoch. 
This does not mean that the kinks produced at present make 
dominant contribution to the energy of  GWs. 
We see above that for given frequency $\omega$, 
the kinks which dominantly contribute to $\omega \frac{dP}{d\omega}$ are 
determined by Eqs.~(\ref{psimaxmat1}),(\ref{psimaxrad}) or 
(\ref{psimaxmat2}). 
Accordingly high frequency modes arise from dense, blunt and old kinks, 
and low frequency modes arise from thin, sharp and new ones. 
The first line of Eq.~(\ref{Omegagw}) corresponds to GWs from new kinks which 
were born after the matter-radiation equality, the second line corresponds 
to GWs from old kinks which were born in the radiation era, 
the third line corresponds to GWs from older kinks produced between the end 
of the inflation and the start of the radiation era and 
the last line corresponds to GWs which came from the most abundant 
and oldest kinks. 
Figure~\ref{fig:spectrum} sketches the shape of the spectrum. 
The spectrum has three inflection points at 
$\omega_1, \omega_2$ and $\omega_3$. 
This is due to change of the type of kinks which mainly contribute to 
$\Omega_{\rm gw}$. 
Positions of these inflection points depend on the reheating 
temperature $T_r$.
If $T_r$ takes the value around its lower bound, say, 
$10$~MeV~\cite{Kawasaki:1999na}, the second inflection point falls in the 
observable region, as we will see.
Even if $T_r$ is so low, the natural value of $H_*$ makes the third inflection 
far above the observable region. 
If the reheating ends immediately and $H_*$ is as small as possible,
the region between the second and third inflection is so short that 
the third inflection enters the observable region.

This is a crude estimation, and in order to derive the realistic spectrum 
we should take into account subtlety described in~\cite{Damour:2001bk,Damour:2004kw} 
where the authors claimed that GWs from kinks are burst-like and hence
GW bursts with rare event rate  (``isolated'' GWs ) should not be counted as 
constituent of the stochastic GW background.
We should calculate the stochastic GW background spectrum using the following 
formulae~\cite{Damour:2001bk,Damour:2004kw}
\begin{equation}
	\Omega_{\rm gw}(f) \sim \frac{3\pi^2}{2}(f t_0)^2h_{\rm conf}^2(f) 
	\label{OmegagwDV}
\end{equation}
\begin{equation}
	h_{\rm conf}^2(f)=\int \frac{dz}{z}\theta(n(f,z)-1) n(f,z)h^2(f,z) 
	\label{hconf}
\end{equation}
\be
n(f,z)=\frac{1}{f}\frac{d\dot{N}}{d\ln z} \label{n}
\ee
\be
	d\dot{N}\sim \frac{1}{4}\theta_m(f,z)(1+z)^{-1}
	\psi_{\rm max}(\omega_z,t)\tilde{N}(\psi_{\rm max}(\omega_z,t),z)
	t^{-1}(z)dV(z), 
	\label{dN}
\ee
\begin{equation} 
	h(f,z)=\frac{G\mu [\psi_{\rm max}(\omega_z,t)]^{1/2}t(z)}{[(1+z)ft(z)]^{2/3}}
	\frac{1+z}{t_0z}\theta(1-\theta_m(f,z)), \label{h}
\end{equation}
\begin{equation}
	\theta_m(f,z) = [(1+z)f t(z)]^{-1/3}, \label{thetam}
\end{equation}
\be
	dV=
	\bc
	54\pi t_0^3 \sqrt{\frac{\Omega_m}{\Omega_r}\frac{T_0}{T_r}}(1+z)^{-9/2}dz 
	& (\mathrm{1st \ matter \ era})\\
	72\pi t_0^3\sqrt{\frac{\Omega_m}{\Omega_r}}(1+z)^{-5}dz 
	& (\mathrm{radiation \ era})\\
	54\pi t_0^3((1+z)^{1/2}-1)^2(1+z)^{-11/2}dz 
	& (\mathrm{2nd \ matter \ era})
	\ec 
	\label{dV}
\ee
\be
	t(z)=
	\bc
	\sqrt{\frac{\Omega_m}{\Omega_r}\frac{T_0}{T_r}}(1+z)^{-3/2}t_0 
	& (\mathrm{1st \ matter \ era})\\
	\sqrt{\frac{\Omega_m}{\Omega_r}}(1+z)^{-2}t_0 & (\mathrm{radiation \ era}) \\
	(1+z)^{-3/2}t_0 & (\mathrm{2nd \ matter \ era})
	\ec
	\label{tz}
\ee
where $f=\omega/2\pi$ and $\omega_z = \omega (1+z)$.
$dV$ means the proper spatial volume between redshifts $z$ and $z+dz$. 
$t(z)$ represents the cosmic time at the redshift $z$. 
$n(\omega,z)$ represents the number of GW bursts at redshift $\sim z$ 
with frequencies $\omega$ superposed in a period 
of $\sim \omega^{-1}$. 
$\tilde{N}(\psi,z)dVd\psi$ is the number of kinks with sharpness 
$\psi \sim \psi +d\psi$ 
in the volume $dV$ at redshift $z$, so 
$\tilde{N}(\psi,z)t^2(z) \sim \frac{dN}{d\psi}(t(z),\psi)$. 
Isolated GW bursts are excluded from the calculation by inserting the step function 
in the integral in Eq.~(\ref{hconf}). 
$h(f,z)$ is the logarithmic Fourier component of the waveform of the GW burst 
from one kink located at redshift $z$ and can contribute 
to GW with frequency $f$. 

\begin{figure}[t]
\begin{center}
\includegraphics[width=15cm]{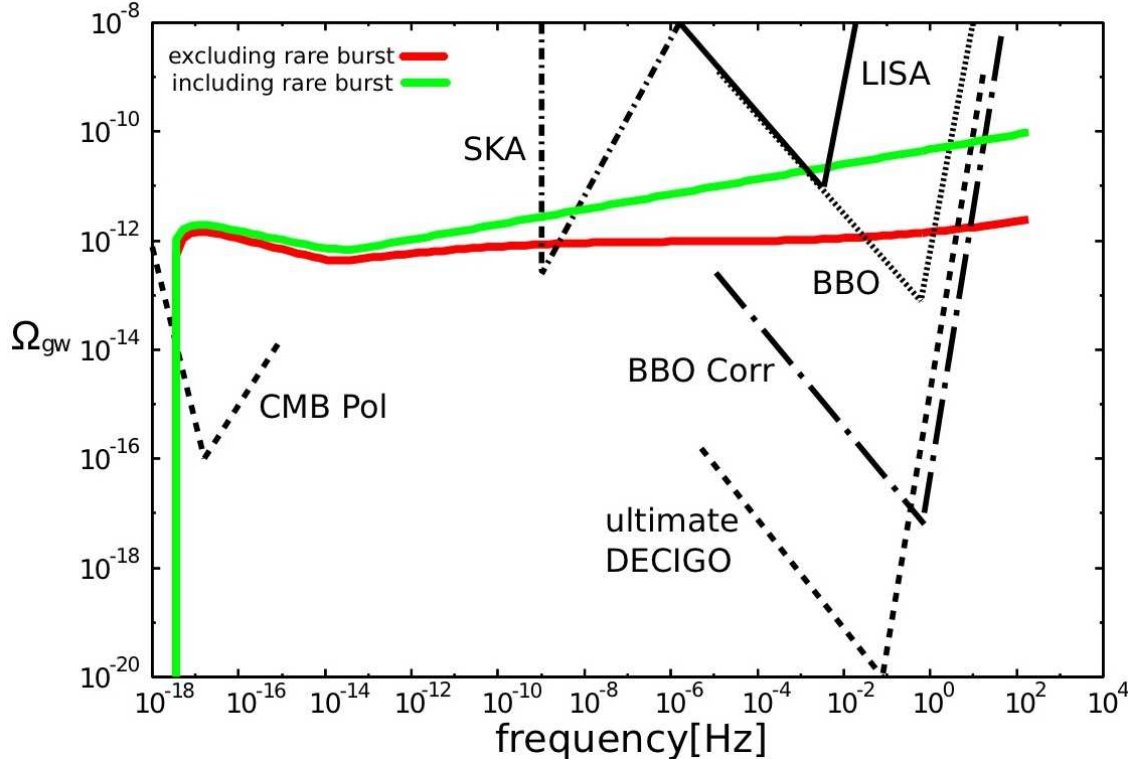}
\caption{$\Omega_{\rm gw}$ in the case where strings emerge 
at the phase transition in the radiation era 
for $G\mu=10^{-7}$ and $T_*\sim 10^{12}$~GeV. 
The upper line represents the estimate using Eq.~(\ref{Omegagw}), i.e. 
including ``rare bursts'', 
and the lower line represents the estimate using Eqs.~(\ref{OmegagwDV})-
(\ref{thetam}), 
i.e. excluding ``rare bursts''. Sensitivity curves of various experiments are shown. 
That of DECIGO is derived from~\cite{Seto:2001qf}. 
That of BBO correlated is derived from~\cite{Buonanno:2004tp}. 
Others are derived from~\cite{Smith:2005mm}.
\label{Omegazuhigh}}
\end{center}

\end{figure}

\begin{figure}[t]
\begin{center}
\includegraphics[width=15cm]{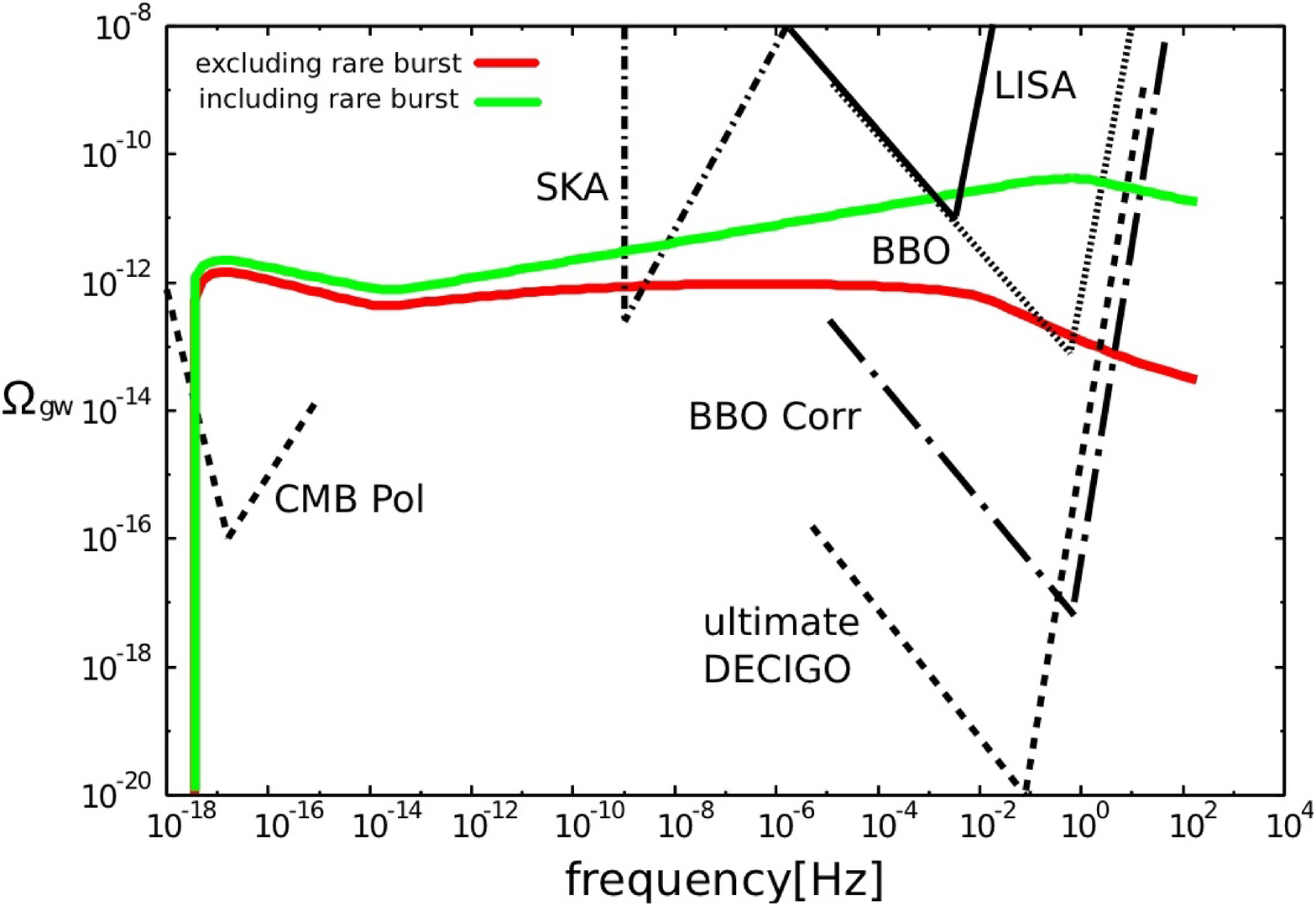}
\caption{$\Omega_{\rm gw}$ in the case where strings emerge at the end of inflation
for $G\mu=10^{-7}$, $T_r\sim 10$~MeV. 
The upper line represents the estimate including ``rare bursts'', 
and the lower line represents the estimate excluding ``rare bursts''.
\label{Omegazulow}}
\end{center}
\end{figure}

The results of calculation are shown in Figure~\ref{Omegazuhigh} and 
Figure~\ref{Omegazulow}. 
We take $G\mu\sim10^{-7}$, close to the current upper bound from CMB 
observation~\cite{Wyman:2005tu}, and show the spectrum with frequency 
$\omega$ from the band of CMB experiments  to that of ground-based GW detectors. 
These figures include both our crude estimate  (\ref{Omegagw}) and the 
improved one given by Eqs.~(\ref{OmegagwDV})-(\ref{tz}).  
We see that the latter is much smaller than the former. 
This is because the spectrum is dominated by GWs emitted recently, 
and recent GW bursts have a more tendency to be isolated. 
The fact that the difference between the two estimates becomes larger 
in higher frequency band might disagree with intuition, 
since the kinks corresponding to high frequency GWs are more abundant. 
However, the higher frequencies of GW are, the smaller the possibility 
that they overlap, because the period of oscillation becomes shorter and the 
extent of the GW beam becomes narrower. 
As a result, higher frequency GWs are more likely isolated in time.
 
%%%%%%%%%%%%%%%%%%%%%%%%%%%%%%%%%%%%%%%%%%%%%%
%%%%%%%%%%%%%%%%%%%%%%%%%%%%%%%%%%%%%%%%%%%%%%
%%%%%%%%%%%%%%%%%%%%%%%%%%%%%%%%%%%%%%%%%%%%%%

In Figure~\ref{Omegazuhigh} we assume that cosmic strings were born by SSB 
at GUT scale and first kinks appeared at the temperature $10^{12}~$GeV (\ref{Tefd}). 
In Figure~\ref{Omegazulow} we assume that strings were produced at the 
end of the inflation with extremely low reheating temperature ($T_r=10$~MeV).
This assumption makes the second inflection point of the spectrum visible 
in the observable frequency band. 
We also assume the inflation energy scale is sufficiently high so that third 
inflection point on the spectrum is far from the observable region.
If we could observe the second inflection, we can deduce the reheating temperature.

Note that the spectrum depends on $G\mu$ via the overall factor $(G\mu)^2$. 
Therefore, when we vary the value on $G\mu$, the spectrum only moves upward or 
downward, and its shape (e.g., the position of bending) does not change. 
This situation is different from the case of GWs from cosmic string loops.
In the case of cosmic string loops,
since different values of $G\mu$ give different values of lifetime of loops, 
the resultant spectral shape is different~\cite{Caldwell:1991jj,DePies:2007bm}.

Let us discuss detectability of this GW background.
We have to see whether GWs from kinks  exceeds not only thresholds of various 
experiments but also GWs from other source.
The spectrum of GWs from loops was discussed in~\cite{Damour:2000wa,Damour:2001bk,Damour:2004kw,Siemens:2006yp,Caldwell:1991jj,DePies:2007bm}, 
and the contribution of loops to GW background is much larger 
than that of kinks if they coexist in some frequency band. 
However, a loop cannot emit GWs with frequencies smaller than the inverse of 
its size. 
Thus there is a cut on the low frequency side of the spectrum 
of GWs from loops 
corresponding to the inverse of the loop size $\sim (\alpha t)^{-1}$. 
If $\alpha \sim G\mu$, the spectrum of GWs from loops begins to appear 
at $\omega \gtrsim 10^{-12}$~Hz,
and this covers the frequency band where both pulsar timing arrays 
and GW detectors have good sensitivity. 
However, $\alpha$ is one of the most unknown parameters in 
the cosmic string model.  
According to some recent simulations~\cite{Ringeval:2005kr}, 
$\alpha$ may be much greater and the broader region may be covered by loops' GW. 
On the other hand, some recent studies~\cite{Siemens:2002dj,Polchinski:2006ee}
show the possibility that $\alpha$ is extremely small, 
say, $\alpha\sim (G\mu)^n$ with $n\gtrsim 1$. 
In such a case, GWs from loops dominate only very high-frequency region and
GWs from kinks may be observable at low-frequency region.
For example, if $\alpha \lesssim 10^{-9}$, 
the band of SKA~\cite{Kramer:2004rwa} can be used for detection of GWs from kinks, 
and for $G\mu \sim 10^{-7}$, $\Omega_{\rm gw}$ reaches the sensitivity of SKA. 
In a more extreme case $\alpha \lesssim 10^{-13}$, 
the sensitivity band of space-borne detectors, 
BBO~\cite{Crowder:2005nr} and DECIGO~\cite{Seto:2001qf} are open for detection of 
kink-induced GWs,
and $\Omega_{\rm gw}$ exceeds the sensitivity of correlated analysis of BBO
for $G\mu \sim 10^{-7}$, 
and that of ultimate-DECIGO for $G\mu \gtrsim 10^{-9}$.\footnote{%%
Note that stochastic GWs from astrophysical sources such as white-dwarf binaries
make a dominant contribution for $\omega \lesssim 1~$Hz~\cite{Farmer:2003pa}
and the observable frequency range is somewhat limited.}%%
In even more extreme case, $\alpha \lesssim 10^{-16}$, 
it may be possible to detect the inflection point in Figure 4 
for $G\mu \gtrsim 10^{-10}$ and determine the reheating temperature as 
$T_r \sim 10$~MeV.

Moreover, GWs from kinks may be detected through CMB observations.
As opposed to GWs from loops, kinks can emit GWs with wavelength 
comparable to the horizon scale.
These GWs induce B-mode polarizations, which is a target of on-going and 
future CMB surveys.
The spectrum of GWs from kinks is quite different from inflationary GWs
and hence its effect on CMB is also expected to be distinguished from that 
of inflationary origin. We will study this issue elsewhere.

%%%%%%%%%%%%%%%%%%%%%%%%%%%%%%%%%%%%%%%%%%%%%%%%%%%%%
\section{Conclusions}   \label{sec:conc}
%%%%%%%%%%%%%%%%%%%%%%%%%%%%%%%%%%%%%%%%%%%%%%%%%%%%%

In this paper, we have considered gravitational waves emitted by kinks on 
infinite cosmic strings. 
We have calculated the spectrum of the stochastic background of such 
gravitational waves and discussed their detectability by pulsar timing 
experiments and space-borne detectors. 
It is found that if the size of cosmic string loops is much smaller than 
that of Hubble horizon, some frequency bands are open for detection of GWs 
originating from kinks.
It can be detected by pulsar timing experiments for $G\mu \gtrsim 10^{-7}$,
and by space-borne gravitational wave detectors for much smaller $G\mu$,
although the latter may be hidden by the loop contribution unless 
the typical loop size is extremely small.
If it is detected, it will provide information on the physics of the early 
Universe, such as phase transition and inflation models.
Moreover, the spectrum shape depends on the thermal history of the Universe,
and hence GWs from cosmic strings can be used as a direct probe into the early 
evolution of the Universe.
Notice that the inflationary GWs also carry information on the thermal history of the Universe
~\cite{Seto:2003kc,Boyle:2005se,Nakayama:2008ip}.
Although the inflationary GWs are completely hindered by GWs from cosmic strings
if the value of $G\mu$ is sizable and GWs from kinks come to dominate in 
low-frequency region,
GWs from cosmic strings also have rich information on the physics of the early Universe.

%%%%%%%%%%%%%%%%%%%%%%%%%%%%%%%%%%%%%%%%%%%%%%%%%%%%%
\appendix 
\section{}\label{sec:app0}
%%%%%%%%%%%%%%%%%%%%%%%%%%%%%%%%%%%%%%%%%%%%%%%%%%%%%

 Here we derive the expression of $dN/d\psi$ given in section 3.
 First, we change the variable from $\psi$ to $\psi_i=\psi(t/t_i)^{2\zeta}$ in Eq.~(\ref{Neq}), 
 where $t_i$ is the time when we set the initial condition.
 Then Eq.~(\ref{Neq}) becomes
 \be
 t\dot{N}(\psi_i,t)+\left(\frac{\bar{\eta}}{\gamma}-2\zeta\right)N(\psi_i,t)=\frac{\bar{\Delta}V}{\gamma^4t^3}g\left(\left(\frac{t_i}{t}\right)^{2\zeta}\psi_i \right),
 \ee
 where the dot now denotes the time derivative at constant $\psi_i$.
 This equation can be easily integrated to obtain
 \be
 \frac{N(\psi,t)}{V(t)}=\frac{\bar{\Delta}}{\gamma^4t^{3-\beta}}\int^t_{\mathrm{max}(t_i,\psi^{1/2\zeta}t)}
 \frac{dt^{\prime}}{t^{\prime 1+\beta}}g\left(\left(\frac{t}{t^{\prime}}\right)^{2\zeta}\psi\right)
 +\left(\frac{t_i}{t}\right)^{3-\beta}\frac{N\left( (t/t_i)^{2\zeta}\psi,t_i\right)}{V(t_i)}. \label{gensol}
 \ee
For the distribution function during the first matter era, we set the initial condition at $t_i=t_*$ 
as $N(\psi,t_*)=0$. Then Eq.~(\ref{gensol}) becomes
\be
\frac{N(\psi,t)}{V(t)}=\frac{\bar{\Delta}_m}{\gamma_m^4t^{3-\beta_m}}\int^t_{\mathrm{max}(t_*,\psi^{1/2\zeta_m}t)}
 \frac{dt^{\prime}}{t^{\prime 1+\beta_m}}g\left(\left(\frac{t}{t^{\prime}}\right)^{2\zeta_m}\psi\right). \label{sol1}
 \ee
 By substituting Eq.~(\ref{g}) into $g$ in Eq.~(\ref{sol1}), performing the integration and omitting terms 
 except for dominant one,
 we get Eq.~(\ref{bunpueasymat1}).
 For $N/V$ during the radiation era, we set $t_i=t_r$ and get
 \be
 \frac{N(\psi,t)}{V(t)}=\frac{\bar{\Delta}_r}{\gamma_r^4t^{3-\beta_r}}\int^t_{\mathrm{max}(t_r,\psi^{1/2\zeta_r}t)}
 \frac{dt^{\prime}}{t^{\prime 1+\beta_r}}g\left(\left(\frac{t}{t^{\prime}}\right)^{2\zeta_r}\psi\right)
 +\left(\frac{t_r}{t}\right)^{3-\beta_r}\frac{N\left( (t/t_r)^{2\zeta_r}\psi,t_r\right)}{V(t_r)}. \label{sol2}
 \ee
 We use Eq.~(\ref{bunpueasymat1}) for $N(\psi,t_r)$.
 Then Eq.~(\ref{sol2}) simplifies to Eq.~(\ref{bunpueasyrad}) by picking only the dominant term.
 The expression during the second matter era (\ref{bunpueasymat2}) can be obtained in the same way.

%%%%%%%%%%%%%%%%%%%%%%%%%%%%%%%%%%%%%%%%%%%%%%%%%%%%%
%\def\thesection{Appendix}
\section{}\label{sec:appA}
%%%%%%%%%%%%%%%%%%%%%%%%%%%%%%%%%%%%%%%%%%%%%%%%%%%%%

Here, we prove kinks which dominantly contribute to 
GWs with frequency $\omega$ are those which satisfy Eq.~(\ref{kinkcon}), 
and evaluate the integral in Eq.~(\ref{I}). 

First of all, we consider the situation that $\omega$ is so small that 
$\left( \psi \frac{dN}{d\psi}\right)^{-1} = \omega^{-1}$ has solutions.
$a^{\prime i}$ has numerous kinks (discontinuities), from blunt 
ones to sharp ones, 
according to Eqs.~(\ref{bunpueasymat1}), (\ref{bunpueasyrad}) and 
(\ref{bunpueasymat2}). 
Let us consider kinks which satisfy $\left( \psi \frac{dN}{d\psi}\right)^{-1} 
\gtrsim \omega^{-1} \ (\Leftrightarrow \psi \gtrsim \psi_{\rm max})$. 
From now on, we call such kinks ``big'' kinks. 
The interval between two kinks with sharpness $\mathcal O (\psi)$ is
roughly given by $\left( \psi \frac{dN}{d\psi}\right)^{-1}$.
Thus the typical interval of big kinks is about $\omega^{-1}$. 
First, we divide the integration range of Eq.~(\ref{I}) 
into short intervals of length $\sim\omega^{-1}$ around each big kink as
\be
	I^{i}_+(k)=\sum_l I^{i}_{+,l}(k),
\ee
where the integer $l$ labels each big kink and 
$I^{i}_{+,l}(k)$ denotes the contribution to $I^{i}_+(k)$ from 
the $l$-th interval. 
Each interval contains one big kink and numerous ``small'' kinks, 
which satisfy $\left( \psi \frac{dN}{d\psi}\right)^{-1} 
\lesssim \omega^{-1} \ (\Leftrightarrow \psi \lesssim \psi_{\rm max})$. 
Let us assume that in the $l$-th interval $a^{\prime i}$ can 
be decomposed as
\be
	a^{\prime i}(u) = \bar{a}_l^{\prime i}(u) + \delta a^{\prime i}_l(u). 
	\label{aprimekatei}
\ee
where $\bar{a}^{\prime i}(u)$ denotes the smooth function (except one big kink) 
which we can get after averaging contributions of small 
kinks to $a^{\prime i}$, 
and $\delta a^{\prime i}_l$ is the contribution of small kinks. 
$\delta a^{\prime i}_l$ discontinuously jumps at each small kink and 
the width of the jump is $\sim \psi^{1/2}$. 
Its average vanishes ($\langle \delta a_l^{\prime i}\rangle=0$) 
since the jump at each kink takes random values.
Then we get
\be
	a^{i}(u)=\bar{a}_l^{i}(u)+\delta a^{i}_l(u), \label{apbunkatu}
\ee
after the integration of Eq.~(\ref{aprimekatei}).  Then, 
\be
	I^{i}_{+,l}(k)=\int_l du \bar{a}_l^{\prime i} 
	\exp(i\omega(u-\mathbf{n}\cdot\bar{\mathbf{a}}_l(u)-
	\mathbf{n}\cdot\delta\mathbf{a}_l(u))/2) + 
	\int_l du \delta a_l^{\prime i} 
	\exp(i\omega(u-\mathbf{n}\cdot\bar{\mathbf{a}}_l(u)
	-\mathbf{n}\cdot\delta\mathbf{a}_l(u))/2). \label{Il}
\ee
Here the integral is performed over the $l$-th interval. 

We are interested in ensemble averages of products of two of $I_+^i$, 
for example, $\langle |I_+^i|^2 \rangle$. 
It contains mean squares of $I^i_{+,l}$'s and cross terms of different $I^i_{+,l}$'s. 
First, we evaluate mean squares of $I^i_{+,l}$'s. 
$\langle |I^i_{+,l}|^2 \rangle$ contains mean squares of the first and 
second terms in Eq.~(\ref{Il}), 
and the averages of the cross terms between them. 
The latter vanish since $\langle \delta a_l^{\prime i}\rangle=0$. 
In order to estimate the mean square of the first term in Eq.~(\ref{Il}), 
we approximate it as
\be
	\int_l du \bar{a}_l^{\prime i} 
	\exp(i\omega(u-\mathbf{n}\cdot\bar{\mathbf{a}}_l(u)-
	\mathbf{n}\cdot\delta\mathbf{a}_l(u))/2) 
	\simeq \sum_a \bar{a}^{\prime i}_l(u_{l;a})
	e^{i\omega(u_{l;a}-\mathbf{n}\cdot\bar{ \mathbf{a}}(u_{l;a}))/2}
	e^{-i\omega\mathbf{n}\cdot\delta\mathbf{a}(u_{l;a})/2}\Delta u_{l;a}. 
	\label{firstapp}
\ee
Here we write the position of the $a$-th small kink in the $l$-th interval 
as $u_{l;a}$, and $\Delta u_{l;a}=u_{l;a+1}-u_{l;a}$. 
We assume that each interval between two small kinks is so short that 
the integrand can be regarded as constant. 
We want to evaluate the mean square of this quantity,
\be
	 \sum_{a,b}\langle \bar{a}^{\prime i}_l(u_{l;a})
	 \bar{a}^{\prime i}_l(u_{l;b})
	 e^{i\omega(u_{l;a}-\mathbf{n}\cdot\bar{ \mathbf{a}}(u_{l;a}))/2}
	 e^{-i\omega(u_{l;b}-\mathbf{n}\cdot\bar{ \mathbf{a}}(u_{l;b}))/2}\rangle
	 \langle e^{-i\omega\mathbf{n}\cdot\delta\mathbf{a}(u_{l;a})/2}
	 e^{i\omega\mathbf{n}\cdot\delta\mathbf{a}(u_{l;b})/2}\rangle 
	 \Delta u_{l;a}\Delta u_{l;b} 
	 \label{first}
\ee
Here we separate the average related to $\bar{a}^{\prime i}_l, \bar{a}^i_l$ and 
that related to $\delta a^{\prime i}_l,\delta a^i_l$, 
assuming that there is no correlation between small kinks and big kinks. 
In order to evaluate the second parenthesis in Eq.~(\ref{first}),
we decompose $\delta a^i_l$ as
\be
	\delta a_l^{\prime i} = \sum_k F^{(k)i}_l(u), \label{Fdef}
\ee
where $F_l^{(k)i}(u)$ is the contribution of kinks of sharpness $\psi \sim \psi_k$, 
so it has discontinuities at intervals 
$\sim \left( \psi_k \frac{dN}{d\psi}(\psi_k)\right)^{-1}$ 
and between two of them its absolute value $\sim \psi_k^{1/2}$. 
Then, 
\begin{align}
   [\langle |\delta \mathbf{a}_l(u_{l;a})
   -\delta \mathbf{a}_l(u_{l;b})|^2\rangle ]^{1/2} 
   & \sim [\langle |\delta a_l^i(u_{l;a})-\delta a_l^i(u_{l;b})|^2\rangle ]^{1/2} 
   \nonumber \\
   & \sim \left[\left<\left(\sum_k \sum_s F^{(k)i}_l (u^k_{l,s})
   (u^{k}_{l,s+1}-u^{k}_{l,s})\right)^2\right>\right]^{1/2} 
   \nonumber \\
   & \sim \left[\sum_k \sum_s \langle (F^{(k)i}_l (u^{k}_{l,s}))^2 \rangle  
   (u^{k}_{l,s+1}-u^{k}_{l,s})^2 \right]^{1/2} 
   \nonumber \\
   & \sim \left[\sum_k \psi_k \times 
   \left(\psi_k \frac{dN}{d\psi}(\psi_k) \right)^{-2} \times 
   \left(|u_{l;a}-u_{l;b}|/\left(\psi_k \frac{dN}{d\psi}(\psi_k) \right)^{-1}\right)    
   \right]^{1/2} 
   \nonumber \\
   & \sim \left[ \sum_k \psi_k \left(\psi_k \frac{dN}{d\psi}(\psi_k) \right)^{-1} 
   \right]^{1/2} |u_{l;a}-u_{l;b}|^{1/2} \nonumber \\
   & \sim \left[ \int^{\psi_{\rm max}} d\psi \psi^{-1}   
   \left( \frac{dN}{d\psi}(\psi) \right)^{-1}\right]^{1/2} |u_{l;a}-u_{l;b}|^{1/2} 
   \nonumber \\
   & \sim \left( \frac{dN}{d\psi}(\psi_{\rm max}) \right)^{-1/2}
    |u_{l;a}-u_{l;b}|^{1/2}.
   \label{delajijou}
\end{align}
To proceed from RHS of the first line to the second line, 
we regard $F_l^{(k)i}(u)$ as constant in the interval between two small kinks. 
Here $u^k_{l,s}$ denotes the position of $s$-th discontinuity of 
$F_l^{(k)i}(u)$. 
$F_l^{(k)i}(u^k_{l,s})$ can be thought of as a probability variable 
whose average is 0 and whose variance is $\sim \psi_k$. 
We can set $\left\langle F^{(k)i}_l(u^{k}_{l,s})
F^{(k^{\prime})i}_l(u^{k{\prime}}_{l,s^{\prime}})\right\rangle=0$ 
unless $k=k^{\prime},s=s^{\prime}$, 
assuming that different kinks are not correlated. 
This enables the second line to be simplified to the third line. 
Then we substitute $\psi_k$ into $(F^{(k)i}_l(u^{k}_{l,s}))^2$, and 
$\left(\psi_k \frac{dN}{d\psi}(\psi_k) \right)^{-1}$ into 
$u^{k}_{l,s+1}-u^{k}_{l,s}$. 
The third factor in the forth line represents the number of 
small kinks in the interval $(u_{l;a},u_{l;b})$. 
When we proceed from the fifth line to the sixth line, 
we changed the sum $\sum_k$ to the integral 
$\int d(\ln \psi)=\int d\psi \psi^{-1}$. 
Using $\psi_{\rm max} \frac{dN}{d\psi}(\psi_{\rm max}) = \omega$ 
and $|u^l_a-u^l_b|\lesssim \omega^{-1}$, we find
 \be
	(\langle |\delta \mathbf{a}_l(u_{l;a})
	-\delta \mathbf{a}_l(u_{l;b})|^2\rangle )^{1/2} 
	\lesssim \psi_{\rm max}^{1/2}\omega^{-1} \ll \omega^{-1}.
\ee
Therefore,  
$\omega\mathbf{n}\cdot(\delta\mathbf{a}_l(u_{l;a})-\delta\mathbf{a}_i(u_{l;b}))$ 
is much less than unity and 
$\langle e^{i\omega\mathbf{n}\cdot(\delta \mathbf{a}_l(u_{l;b})
-\delta \mathbf{a}_l(u_{l;a}))/2} \rangle$ is $\sim 1$. 
Then Eq.~(\ref{first}) is written as 
\be
	 \sum_{a,b}\langle \bar{a}^{\prime i}_l(u_{l;a})\bar{a}^{\prime i}_l(u_{l;b})
	 e^{i\omega(u_{l;a}-\mathbf{n}\cdot\bar{ \mathbf{a}}(u_{l;a}))/2}
	 e^{-i\omega(u_{l;b}-\mathbf{n}\cdot\bar{ \mathbf{a}}(u_{l;b}))/2}\rangle 
	 \Delta u_{l;a}\Delta u_{l;b} = 
	 \left<\left|\int_l du \bar{a}^{\prime i}_l(u)
	 e^{i\omega(u-\mathbf{n}\cdot\bar{\mathbf{a}}(u))/2}\right|^2 \right>. 
	 \label{first2}
\ee
The result is same as that derived without the contribution from small kinks. 
Then, the RHS of Eq.~(\ref{first2}) can be calculated as Eq.~(\ref{Idisc}), 
and its magnitude is
 \be
	 \left( \frac{\psi_l^{1/2}}{\omega}\right)^2, \label{barjijou}
 \ee
where $\psi_l$ denotes the sharpness of $l$-th big kink.
 
In order to estimate the mean square of the second term in Eq.~(\ref{Il}), 
we approximate it as
\be
	\int_l du \delta a_l^{\prime i} 
	\exp(i\omega(u-\mathbf{n}\cdot\bar{\mathbf{a}}_l(u)
	-\mathbf{n}\cdot\delta\mathbf{a}_l(u))/2) 	
	\simeq \sum_k\sum_s F^{(k)i}_{l}(u^{k}_{l,s})
	\exp(i\delta^k_{l,s})(u^{k}_{l,s+1}-u^{k}_{l,s}), 
	\label{second}
\ee
where 
$\delta^k_{l,s} = \omega(u^{k}_{l,s}-\mathbf{n}
\cdot\bar{\mathbf{a}}_l(u^{k}_{l,s})
-\mathbf{n}\cdot\delta\mathbf{a}_l(u^{k}_{l,s}))/2$. 
The mean square of Eq.~(\ref{second}) is 
\begin{align}
   &\left\langle \left(\sum_k\sum_s F^{(k)i}_{l}(u^{k}_{l,s})
   \exp(i\delta^k_{l,s})(u^{k}_{l,s+1}-u^{k}_{l,s})\right)^2\right\rangle 
   \nonumber \\
   & = \sum_k\sum_s \left<(F^{(k)i}_{l}(u^{k}_{l,s}))^2\right>
   (u^{k}_{l,s+1}-u^{k}_{l,s})^2 \nonumber \\
   &\sim \sum_k \psi_k   \left(\psi_k \frac{dN}{d\psi}(\psi_k) \right)^{-2} 
   \left(\omega^{-1} / \left(\psi_k \frac{dN}{d\psi}(\psi_k) \right)^{-1}\right) 
   \nonumber \\
   &\sim \sum_k \left(\psi_k \frac{dN}{d\psi}(\psi_k) \right)^{-1}\omega^{-1} 
   \nonumber \\
   &\sim \int^{\psi_{\rm max}}d\psi \psi^{-1}
   \left( \frac{dN}{d\psi}(\psi) \right)^{-1}\omega^{-1} 
   \nonumber \\
   &\sim \left( \frac{dN}{d\psi}(\psi_{\rm max}) \right)^{-1}\omega^{-1}.
   \label{second2}
\end{align}
Remembering $\psi_{\rm max}\frac{dN}{d\psi}(\psi_{\rm max})=\omega$ and 
$\psi_{\rm max}<\psi_l$, we find (\ref{second2})$<$(\ref{barjijou}).
Eventually, the mean square of $I^i_{+,l}$ is roughly estimated 
as (\ref{barjijou}). 
In other words, in each interval around each big kink, it is sufficient to 
consider only the isolated big kink, while neglecting small kinks.
 
Next, we consider the cross terms of different $I^i_{+,l}$s, such as 
$\langle I^i_{+,l}I^{i*}_{+,m} \rangle$. This should vanish, and we can 
explicitly check this by straightforward calculation. 
To do so, we divide $I^i_{+,l}$ and $I^{i}_{+,m}$ as Eq.~(\ref{Il}) and evaluate 
the mean squares and the averages of the cross terms of the two term, 
using above approximations, such as Eqs.~(\ref{firstapp}) 
and (\ref{second}).  
 
As a result, the mean square of $I^i_+$ can be evaluated by summing up 
Eq.~(\ref{barjijou}) for each $l$. Then we obtain
 \begin{align}
	\left\langle |I^{i}_+|^2 \right\rangle & \sim \sum_l \psi_l \omega^{-2} 
	\nonumber \\
	& \sim \int^1_{\psi_{\rm max}} d\psi \frac{dN}{d\psi}(\psi)\psi 
	\omega^{-2} \times L\nonumber \\
	& \sim \psi_{\rm max}\frac{dN}{d\psi}(\psi_{\rm max})\times 
	\psi_{\rm max}  \omega^{-2}\times L ,
\end{align}
where $L$ denotes the integration range of $I_+^i$. 
This implies that the greatest contribution to 
$\left\langle |I^{i}_+|^2 \right\rangle$ 
comes from kinks which satisfy $\psi\sim\psi_{\rm max}$. 
Such kinks dominantly contribute to GWs with frequency $\sim\omega$.

So far we have discussed the case where 
$\psi_k \frac{dN}{d\psi}(\psi_k) \sim \omega$ has a solution.
However, if $\omega$ is so large that 
$\psi_k \frac{dN}{d\psi}(\psi_k)  \ll \omega$
is satisfied for arbitrary values of sharpness,
all kinks are thought of  as ``big kinks''. 
Therefore, the contribution from each interval of length $\sim \omega$ 
around each kink becomes (\ref{barjijou}). 
That from regions far from any kinks is exponentially small when 
$\omega \rightarrow \infty$. Eventually,
\begin{align}
	\left\langle |I^{i}_+|^2 \right\rangle 
	& \sim \int^1_0 d\psi \frac{dN}{d\psi}(\psi)\psi \omega^{-2} 
	\times L\nonumber \\
	& \sim \psi_{\rm max}\frac{dN}{d\psi}(\psi_{\rm max})
	\times \psi_{\rm max}  \omega^{-2}\times L, 
\end{align}
where $\psi_{\rm max}$ denotes the value of $\psi$ at which 
$\frac{dN}{d\psi}(\psi)$ has a peak. This implies that kinks which satisfy 
$\psi\sim\psi_{\rm max}$ dominantly contribute to 
$\left\langle |I^{i}_+|^2 \right\rangle$ and GWs of frequency $\sim\omega$.
Thus we have proved the validity of Eq.~(\ref{omedPdome}).

\section{}   \label{sec:appB}
%%%%%%%%%%%%%%%%%%%%%%%%%%%%%%%%%%%%%%%%%%%%%%%%%%%%%

Here we discuss a subtlety related to validity to use the distribution function of 
kinks [Eqs.~(\ref{bunpueasymat1}), (\ref{bunpueasyrad}) 
and (\ref{bunpueasymat2})]. 
These formulae are derived without considering gravitational backreaction. 
The distribution may be altered if such an effect is taken into account. 
It may be necessary to define the residual lifetime for blunt kinks 
and set lower cutoff of sharpness. 
It is difficult to clarify how we should take into account 
this effect at this moment. 
However, at least we can find a crude condition which must be satisfied 
regardless of the detail of backreaction;
the energy of GWs emitted from strings must be less than the string energy.
This condition is expressed as
\be
	\int ^{\omega} d\omega^{\prime} \frac{dP}{d\omega^{\prime}} \bigg|_{\rm tot} \times t < \mu t. 
	\label{consis}
\ee
LHS represents the energy emitted from kinks on one infinite string 
in a Hubble horizon per Hubble time, 
and RHS denotes the energy of one infinite string in a Hubble horizon.  
(Note that this is only a necessary condition that the backreaction does not affect kink distribution.)

First, let us assume that strings emerged in the radiation era. 
In the radiation era, the condition (\ref{consis}) is satisfied for
\begin{equation}
	t<(10G\mu)^{3/E_r}t_*\sim (10G\mu)^{-7.9}t_* ~~~(
	~\Leftrightarrow ~T>(10G\mu)^{-3/2E_r} T_* \sim (10G\mu)^{3.9}T_*) . \label{confortrad}
\end{equation}
($E\equiv 8\zeta-\beta, E_r\simeq -0.38, E_m \simeq -0.4$.) 
For Eq.~(\ref{confortrad}) to be satisfied  in the whole radiation era, 
\be
	 T_*<(10G\mu)^{3/2E_r}T_{eq} \sim (10G\mu)^{-3.9}T_{eq}. 
	 \label{conforT}
\ee
If we take $G\mu \sim 10^{-7}$, this becomes $T_* \lesssim 10^{14}$~GeV. 
In the matter era, the condition (\ref{consis}) is written as
\be
	t >(10G\mu)^{3/E_m}\left( \frac{T_*}{T_{\rm eq}} \right)^{-2E_r
	/E_m}t_{\rm eq} \sim (10G\mu)^{7.5}
	\left( \frac{T_*}{T_{\rm eq}} \right)^{1.9} t_{\rm eq} . 
	\label{confortmat}
\end{equation}
The condition that the backreaction is not problematic in the matter era 
also leads to  (\ref{conforT}). 
Eventually, if we assume that kinks had not appeared until friction domination ended or 
strings emerged at low temperature at which friction can be neglected, 
the gravitational backreaction is not important.

Next, let us assume that strings were born at the end of inflation. 
It is easy to see that (\ref{consis}) is satisfied in the first 
matter era, using Eq.~(\ref{powtotmat1}). 
After the first matter era, the situation depends on 
whether the reheating temperature exceeds $T_c$ [Eq.~(\ref{Tefd})] or not. 
If $T_r < T_c$, there is no period when the friction works 
and kinks produced in the first matter era survive. 
The peak of $\omega\frac{dP}{d\omega}|_{\rm tot}$ consists of 
contribution from kinks produced around the turning point from 
the first matter era to the radiation era.
Therefore, the above discussion applies  and the condition (\ref{consis}) is 
satisfied all the time.  
On the other hand, in the case of $T_r > T_c$, 
the friction becomes problematic in the early stage of the radiation 
dominated era. 
If all kinks disappear in this stage and kinks restart to emerge 
at the end of friction-domination, 
the condition (\ref{consis}) is never violated as discussed above. 
In the opposite case, where all kinks survive the friction-dominated era, 
(\ref{consis}) is not guaranteed. 
In such a case, the largest contribution to 
$\int d\omega \frac{dP}{d\omega}|_{\rm tot}$ comes from kinks produced 
around the end of the first matter era. 
The condition (\ref{consis}) is satisfied if
\be
	10G\mu \left(\frac{T_r}{T_{eq}}\right)^{-2E_r/3}<1.
\ee
For $G\mu=10^{-7}$, this leads to $T_r/M_{\rm pl}\lesssim10^{-4}$.
Therefore the gravitational backreaction might be able to be neglected 
unless the reheating temperature is so high.

%%%%%%%%%%%%%%%%%%%%%%%%%%%%%%%%%%%%%%%%%%%%
\begin{acknowledgments}
%%%%%%%%%%%%%%%%%%%%%%%%%%%%%%%%%%%%%%%%%%%%

K.N. would like to thank the Japan Society for the Promotion of Science for financial support.
This work is supported by Grant-in-Aid for Scientific research from the Ministry of Education,
Science, Sports, and Culture (MEXT), Japan, No.14102004 (M.K.)
and No. 21111006(M.K. and K.N.)
and also by World Premier International
Research Center Initiative (WPI Initiative), MEXT, Japan.

 %%%%%%%%%%%%%%%%%%%%%%%%%%%%%%%%%%%%%%%%%%%%
\end{acknowledgments}
%%%%%%%%%%%%%%%%%%%%%%%%%%%%%%%%%%%%%%%%%%%%

%%%%%%%%%%%%%%%%%%%%%%%%%%%%%%%%%%%%%%%%%%%%

 \end{document}